\documentclass[prd,aps,amsfonts,nofootinbib,longbibliography,notitlepage,twocolumn]{revtex4-1}
\usepackage{graphicx}
\usepackage{xcolor}
\usepackage{rotating}
\usepackage{amsmath,amssymb,graphics,amsthm,isomath}
\usepackage{bbm}
\usepackage{bm}
\usepackage{array}
\newcolumntype{P}[1]{>{\centering\arraybackslash}p{#1}}
\usepackage{multirow}

\usepackage[colorlinks=true, urlcolor=violet, linkcolor=blue, citecolor=red, hyperindex=true, linktocpage=true]{hyperref}
\usepackage[capitalise,compress]{cleveref}
\allowdisplaybreaks

\numberwithin{thm}{section}

\renewcommand{\thesection}{\arabic{section}}
\renewcommand{\thesubsection}{\thesection.\arabic{subsection}}

\makeatletter
\renewcommand{\p@subsection}{}
\renewcommand{\p@subsubsection}{}
\makeatother

\usepackage{xcolor}
\usepackage{mathtools}

\def\bea{\begin{eqnarray}}
\def\eea{\end{eqnarray}}
\def\be{\begin{equation}}
\def\ee{\end{equation}}
\def\bes{\begin{subequations}}
\def\ees{\end{subequations}}
\def\bed{\begin{displaymath}}
\def\eed{\end{displaymath}}
\def\beal{\begin{aligned}}
\def\eeal{\end{aligned}}
\def\bew{\begin{widetext}}
\def\eew{\end{widetext}}
\def\beit{\begin{itemize}}
\def\eeit{\end{itemize}}
\def\bea{\begin{array}}
\def\eea{\end{array}}
\def\been{\begin{enumerate}}
\def\eeen{\end{enumerate}}

\usepackage{dsfont}
\usepackage{docmute}
\renewcommand{\thesection}{\arabic{section}} %COMMAND FOR ARXIV
\renewcommand{\thesubsection}{\thesection.\arabic{subsection}}
\renewcommand{\thesubsection}{\thesubsection.\arabic{subsubsection}}

\begin{document}
\title{Non-equilibrium phase transitions in competitive markets caused by network effects}

\author{Andrew Lucas}
\email{andrew.j.lucas@colorado.edu}
\affiliation{Department of Physics, University of Colorado, Boulder CO 80309, USA}

\date{\today}

\begin{abstract} % abstract
Network effects are the added value derived solely from the popularity of a product in an economic market.  Using agent-based models inspired by statistical physics, we propose a minimal theory of a competitive market for (nearly) indistinguishable goods with demand-side network effects, sold by statistically identical sellers.
With weak network effects, the model reproduces conventional microeconomics: there is a statistical steady state of (nearly) perfect competition. Increasing network effects, we find a phase transition to a robust non-equilibrium phase driven by the spontaneous formation and collapse of fads in the market.  When sellers update prices sufficiently quickly, an emergent monopolist can capture the market and undercut competition, leading to a symmetry- and ergodicity-breaking transition.   The non-equilibrium phase simultaneously exhibits three empirically established phenomena not contained in the standard theory of competitive markets: spontaneous price fluctuations, persistent seller profits, and broad distributions of firm market shares.
\end{abstract}

\maketitle

\section{Introduction}

Economists have an established theory of supply and demand for highly competitive markets in equilibrium.  However, our everyday life is full of markets \emph{far from static equilibrium}, with price fluctuations \cite{malin} and fad-driven dynamics \cite{salganik, aral}.
A standard assumption is that this is caused by external shocks \cite{acemoglu, carvalho} or technological growth \cite{baumol}, but that the market (if ``left alone") would equilibrate.  Persistent dynamics can arise from market friction or ``long-term" strategizing \cite{grandmont}.

To understand the robustness of these assumptions, we devise a theory of a market for a single good, with  $M$ competitive sellers and $N$ buyers, using agent-based models inspired by statistical physics \cite{galam91, galam97, jpb2005, jpb2007,yanagita}; see \cite{bouchaudreview} for a review.    The  ingredients in these models which are natural from a physics viewpoint are heterogeneity among $N$ buyers (whose individual preferences among the sellers vary), and network effects \cite{granovetter, durlauf, dube,grilo}: the preference of buyers to select a product which is already popular.  (This does \emph{not} refer to granularity of a ``social  network" in which buyers interact.)  In the physics context, we model the market using a time-dependent Potts model, where buyers correspond to ``spins", with the seller they buy from given by their ``spin state".  Heterogeneous preferences are  random fields, and network effects are all-to-all (mean-field) ferromagnetic interactions.   The ground state of the Potts model gives a (possibly multi-valued) high-dimensional ``demand curve" \cite{leelucas}.  Sellers individually profit-maximize in this high-dimensional landscape, adjusting prices (corresponding to time-dependent uniform fields in the Potts model) with time.  Buyers adjust choices in response and we numerically simulate the dynamics.

When network effects (i.e. spin interactions) are very weak, there is (nearly) perfect competition: a statistical steady state with negligible seller profits.  As network effects increase, there is a non-equilibrium phase transition, after which fads spontaneously form \cite{leelucas}.   Sellers exploit this condensation of buyers onto their good by raising their price, after which other sellers undercut them;  this cycle causes \emph{persistent dynamics}, reminiscent of idiosyncratic price fluctuations observed in many markets \cite{malin}.  Sellers make finite time-averaged profit in this non-equilibrium phase, as we simultaneously observe numerically a broad distribution of sellers' market shares.  As heterogeneity vanishes or buyer dynamics become sufficiently slow, it is also possible for one seller to permanently capture the market and price out possible competitors.  Hence we find two transitions: from an equilibrium symmetric phase to a non-equilibrium phase which breaks the permutation symmetry (all sellers are just as likely to be preferred) at any fixed time, but not after time-averaging;  then to a non-ergodic and symmetry-broken phase where a monopolist captures the market.  

{\color{black}Our work complements a large body of recent work investigating how network effects can disrupt the simple model of competitive equilibrium.  For example, network effects cause just two competing firms to change pricing strategies \cite{grilo}, even leading to temporal dynamics \cite{cabral,jdg16}. Work has been more limited on markets with many firms, where the focus has been on classifying multiple equilibria due to network effects \cite{jpb2007,leelucas}.  But with non-dynamical sellers, a market with many firms will ultimately reach static equilibrium even with network effects \cite{leelucas}.  The punchline of this paper is that competing profit-maximizing sellers can generically destabilize this static picture. With strong network effects, there is a robust phase in which large and unpredictable temporal fluctuations of market shares persist to infinite time. This phenomenon is endogenous (not caused by external shocks), yet still \emph{preferable} to the sellers, who make far higher profits.  Sellers could therefore prefer to operate in markets far from equilibrium, characterized by persistent boom-and-bust cycles. Our model thus proposes how some highly competitive markets with many interchangeable goods might fail to equilibrate, even in the long run.}

\section{Model Setup}

We now introduce the details of our model.   We do not attempt to capture all possible complex features of economics; rather, we focus on a minimal model which can realize the phenomenology outlined above.  Buyers are labeled with $\alpha=1,\ldots, N$, and sellers with $i=1,\ldots,M$.   Time is labeled in integer steps: $t=1,2,\ldots$.   At each time step, seller $i$ sells a good at uniform price $p_i$ to the entire market.   Let $q_i(t)$ denote the fraction of buyers who select good $i$ at time $t$.  We assume that a buyer can always purchase from their desired seller, and can also choose not to buy from any seller.

When we increment $t$ by 1, first each of the $N$ buyers updates their decision with probability $\rho$.  Buyer $\alpha$ picks their next decision by maximizing the utility $U_{\alpha,i}$ of choice $i$ at time $t$, which we model by \cite{leelucas} \begin{equation}
U_{\alpha,i}(t) = u_{\alpha,i} + Jq_i(t-1) - p_i(t-1). \label{eq:Ueq}
\end{equation}
where the random fields $u_{\alpha,i}$ gives intrinsic heterogeneity in buyers' choices, the $+Jq_i$ term models network effects (with $J\ge0$ denoting their strength), and $-p_i$ denotes the loss of utility from paying more for a good.  We take $u_{\alpha,i}$ to be independent and identically distributed Gaussian random variables with mean $\mu$ and variance $\sigma^2$.   
% drawn from cumulative distribution function  \begin{equation}
%F(x) = \mathbb{P}(u_{\alpha,i} \le x).
%\end{equation}
If buyer $\alpha$ updates, they buy from seller $i$ if $U_{\alpha,i} \ge U_{\alpha, j}$ for all $j$, and from no seller ($x=0$) if all $U_{\alpha,i}<0$.    We can include this last effect by simply defining $U_{\alpha,0}=0$ for all buyers.
%Numerically, we repeatedly query the buyers $\alpha$ who flip at time $t$ until all of them are satisfied (as they may wish to change their choice based on another buyer's update); this process will converge since buyers are cooperative.

It is then the sellers' turn to act.  One seller $i$,  chosen uniformly at random, will update their price $p_i$.   To do this, the seller follows ``textbook economics":  they query each buyer $\alpha$ and determine the price point \begin{equation}
p^*_{\alpha,i} = p_i + U_{\alpha,i} - \max_{0\le j \le N, j\ne i} U_{\alpha,j}
\end{equation}
 at which that buyer would be willing to purchase their good (which might be negative!). They then set their price by maximizing profit {\color{black} (per total buyer, implicit henceforth)} \begin{equation}
  \pi_i(p_i) = \frac{p_i}{N}\sum_{\alpha=1}^N  \mathrm{\Theta}(p_{\alpha,i}^*-p_i).
 \end{equation}
 {\color{black} If $p_i^*$ is the value at which $\pi_i$ above is maximized, sellers choose $p_i \rightarrow \max(0,p_i^*)$.}
 
 {\color{black} We assume for now that the sellers have no production costs.  Inclusion of production costs into the model does not change the nature of the phase diagram discussed below: see the Supplementary Information (SI).  The model free of production costs may be more suited as a model of markets such as art, fashion, software, or entertainment.  In particular, in both software \cite{gallaugher,corts} and fashion \cite{yamaguchi}, there is empirical evidence of strong network effects.  Fashion markets are notorious for exhibiting persistent dynamics \cite{sproles}, and thus may be an ideal setting for empirical tests of this theory. Alternatively, a seller selling digitally-downloaded software can instantly supply an arbitrary number of goods to interested buyers, analogously to the dynamics in our model.
 
Note that in this simplified toy model, buyers must re-buy a good in each time step (though wait, on average, $\rho^{-1}$ time steps before adjusting their utility maximization calculation).  One can imagine that these goods are perishable, and/or there is a subscription model for purchasing, such that buyers consistently re-buy goods at each time. Sellers can and must sell to any buyer who wants to purchase from them, suggesting that sellers can and will produce goods on demand, as in the digital download market, or perhaps even ``fast fashion" markets \cite{cachon}.  However, we emphasize that the model's phase diagram is not sensitive to each precise detail. The core phenomena in this model are likely present in alternative microscopic models with the same key features: incorporating network effects among buyers making discrete (rather than continuous) purchases from competing sellers.  With this perspective in mind, we proceed with the analysis.}
% It is straightforward to account for sellers' production costs in this model (see SI), but this generalization does not change the phase diagram.
% We will see that the seller's determination of the profit-maximizing price can thus be dangerously wrong.
%  The sellers \emph{do not} incorporate network effects into their calculation; they simply query how much excess utility their customers have, and at what price their non-customers would be willing to buy at.    Sellers can (and will) make large changes to $p_i$ in a single time step if it maximizes their expected $\pi_i$.
%    However, we assume that sellers are not capable of predicting how network effects cause a further chain reaction of buyer decisions that follow from changing $q_i$.
%To help sellers with $q_i=0$ to ``reset", when numerically simulating our model, we add a very small amount of random noise to the seller's estimate of $\pi_i(p_i)$.

\section{Phase Diagram of the Model}

In a naively rational marketplace, all sellers should perform just as well as all others; there is no objectively better good.   In the language of statistical physics, there should be an \emph{emergent} $\mathrm{S}_M$ permutation symmetry in the model in the $N\rightarrow \infty$ limit, for each random realization of $u_{\alpha,i}$.  {\color{black} Note that any given realization of randomness ($U_{\alpha,i}$) microscopically breaks permutation symmetry.  However, as is common in statistical physics, we consider this effect to be meaningful only if it has macroscopic effects: namely, $\lim_{N\rightarrow \infty} q_i \ne M^{-1}$.  When this identity holds, we say that there is permutation symmetry breaking; when $\lim_{N\rightarrow \infty} q_i = M^{-1}$, we call the phase permutation-symmetric, since we could permute the sellers' labels $i$ without changing the macroscopic observable $q_i$ (at large $N$).

In a dynamical rational marketplace, we should expect that by the definition above, the market has permutation symmetry among the $M$ sellers: buyers can instantaneously, and without taking on added costs, switch buying from seller $i$ to $j$.  Moreover, sellers can and will make instantaneous and large changes to their prices to undercut other sellers, when possible.  In these circumstances, we should find perfect competition, where all sellers have equal market shares and make no profit as $\sigma \rightarrow 0$ (see Appendix \ref{app:PC}).  After all, without network effects, if one seller makes profit at price $p_0$, another seller can undercut them with price $p_0 - \delta$, with $\delta \rightarrow 0^+$, thus stealing their entire market share. Prices $p_i\rightarrow 0$ over time, so in the long run ($t\rightarrow\infty$) sellers make no profit.  At $\sigma=0$, this effect is not sensitive to whether $N$ is large or small.
}

% It was shown in \cite{} how to efficiently determine $q_i$ when $N\gg 1$.  Define the function \begin{equation}
% G = \int\limits_0^\infty \mathrm{d}z  \left(-1+\prod_{i=1}^M F(z-Jq_i+p_i)\right).
% \end{equation}
% Then in equilibrium, \begin{equation}
% q_i = \frac{\partial G}{\partial p_i}, \label{eq:demand}
% \end{equation}
%which is readily confirmed by noting that the integrand of $\partial G/\partial p_i$ is the conditional probability that a buyer has maximal utility to buy good $i$ subject to $u_{\alpha,i}=z$.  The key feature of the random-field Potts models is the generic presence of \emph{phase transitions}, where (\eqref{eq:demand}) has \emph{multiple solutions} $q(p)$.   We interpret this as the fact that the demand curve (or more generally, surface) is multi-valued.  The existence of these multiple branches is key in what follows.

%Clearly there are numerous assumptions in the above model that can be relaxed (and whose relaxation is likely interesting).  Our goal is just to present a minimal model for a non-equilibrating marketplace.

\begin{figure}[t]
\includegraphics[width=\columnwidth]{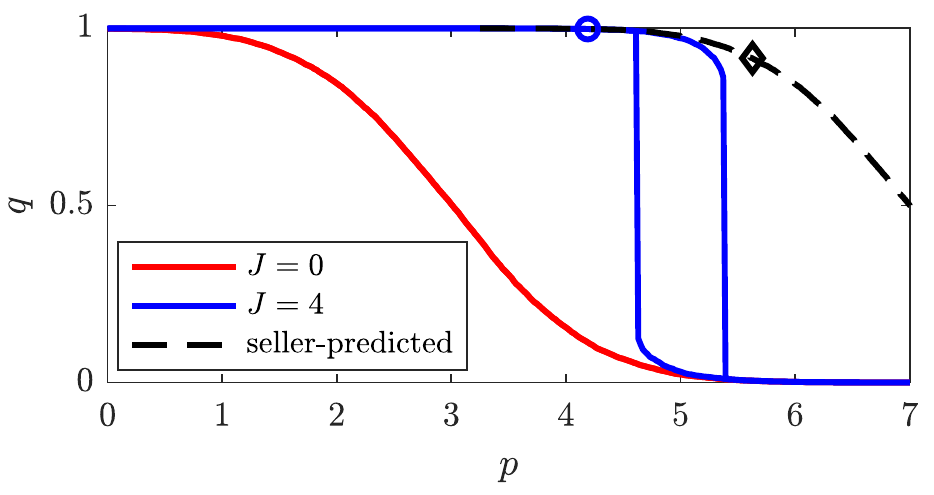}
\caption{The supply and demand problem faced by a single seller.   We take $\mu=3$ and $\sigma=1$.   At the blue circled point on the true demand curve, the seller (neglecting that network effects allow their higher price point) predicts the demand curve is the black dashed line, and the profit maximizing point is at the black diamond, which will lead to a market crash.}
\label{fig:mono}
\end{figure}

However, network effects can drastically modify this picture.  It is useful to review the profit maximization problem faced by a single seller ($M=1$) in a market with network effects \cite{nadal}.  Figure \ref{fig:mono} shows $q(p)$ and $\pi(p)$ for both small and large $J$.  When $J$ is large, the demand curve $q(p)$ is multi-valued: this is a hysteresis loop well-known from the phase diagram of a ferromagnet in statistical physics, where it is possible for two different collective behaviors to be stable (in this case, many or few buyers purchasing the good).   As an extreme limiting case, if $\sigma \rightarrow 0$ but $J$ is finite, then buyers are willing to buy at price $p=\mu$ (the mean of $u_{\alpha,i}$) when $q=0)$, but at $p=\mu+J$ when $q=1$; hence for $\mu < p < \mu +J$ there are two possible (stable) market outcomes.  At finite $\sigma$, it turns out that the monopolist's preferred price -- namely, the price which maximizes $\pi(p) = q(p)p$, is very close to the critical price $p$ at which the upper branch of $q(p)$ ceases to exist.  If $p$ is raised beyond this point, the market will ``crash" (there is a discontinuous phase transition where buyers no longer purchase from this seller).   In our simple model, where the seller does not anticipate network effects, they will tend to overprice their good and cause a market crash, as shown in Figure \ref{fig:mono}.

%It may be the case that the $\pi$-maximizing price $p$ changes.   When $\eta>0$, the seller will slowly adjust their price to maximize $\Pi$.   The optimization problem can be solved using variational calculus and mapped on to a one-dimensional particle of mass $m=1/2\eta$ moving in potential energy $-\pi(p)$ as a function of position $p$.  For $T\rightarrow \infty$, the $\Pi$-maximizing trajectory should be optimized over all final positions $p(T)$ and will correspond to $|p(t)-p_*|$ decaying exponentially quickly at sufficiently large $T$.  However if the period of time $T$ is sufficiently short, the monopolist will prefer a trajectory which stays farther from $p_*$ to maximize the short-term integrated profit $\Pi$.  These observations are simple for one seller, but will have interesting consequences once there are multiple sellers in the marketplace.

What happens if the monopolist has to compete?   {\color{black} We have already noted that in the absence of network effects, $M\ge 2$ sellers will simply undercut each others' prices for greater market share, pushing $p_i\rightarrow 0$ if $\sigma=0$.  If $\sigma>0$, sellers stop undercutting each other when} $p\sim \sigma (\log \frac{M}{2})^{-1/2}$, which is the price at which a seller can expect to keep a fraction of their buyers even if all other sellers set $p=0$: see SI.  The $(\log M)^{-1/2}$ scaling comes from Gaussianity of $u_{\alpha,i}$ and is not universal.   The resulting market will then enter a statistically steady state where sellers continue to make small price adjustments to capture a handful of marginal sellers, but they each have average market share $q_i \sim M^{-1}$, up to subleading corrections in $1/N$.  We conclude that this phase is (nearly) perfect competition (PC); crucial features of this phase are low seller profits, statistical stationarity (no macroscopic dynamics in time), and emergent permutation symmetry.  An order parameter is \begin{equation}
\frac{1}{Q} = \frac{1}{(1-q_0)^2}\sum_{i=1}^M q_i^2.
\end{equation}
$Q$ is the ``effective number of sellers" if the market share were equally distributed at any time $t$; here $q_0$ denotes the fraction of buyers who exited the market; in the PC phase, $Q= M$ (when $N\rightarrow \infty$).  %In simulations shown in this paper, we choose $\mu\gg \sigma,J$ which means that $q_0 \approx 0$.
%In the SI, we confirm this phenomenology in numerical simulations, along with the vanishing of fluctuations in $q_i$ as $N\rightarrow \infty$.  [DO THIS.  DO FLUCTUATIONS GO $N^{-1/3}$?  ATYPICAL...]

Suppose now that we consider the opposite limit where $J $ is finite while $\sigma=0$.  Now an emergent monopolist captures the market and sets $p=J - 0^+$, crowding out any other seller.   This trivial limit spontaneously breaks the permutation symmetry group $\mathrm{S}_M$, since buyers collectively and ``irrationally" choose a single seller to buy from, even though that seller is no better than any other.  $Q\sim1$ for the symmetry-broken (SB) phase.

\begin{figure*}[t]
\includegraphics[width=\textwidth]{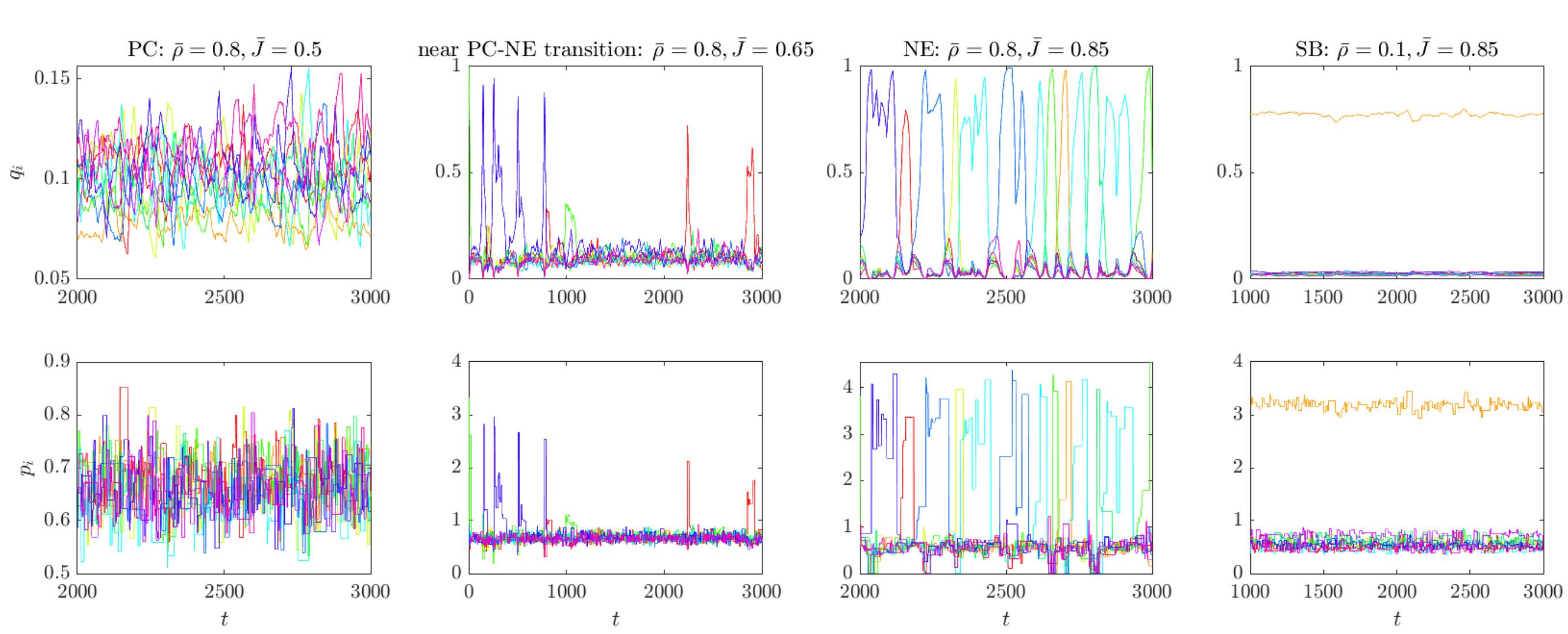}
\caption{Dynamics in a competitive market with $M=10$ sellers (each different color) and $N=5000$ buyers.  The top row plots seller $i$'s market share $q_i(t)$; the bottom row plots price $p_i(t)$.  In the NE phase, the mechanism driving oscillations is clearly visible: the seller with high market share raises price, and will be undercut unless they can catch the effect in time to lower their price and stop the cascade of buyers to another seller.}
\label{fig:dynamic}
\end{figure*}

What happens when $J/\sigma$ is neither 0 nor $\infty$?  When $J$ is sufficiently small, \eqref{eq:Ueq} will only have one solution for fixed prices \cite{leelucas}, and this implies the PC phase is stable.  For sufficiently large $J$, \eqref{eq:Ueq} can have multiple solutions. This occurs when the gain in utility $JN^{-1}$ for good $i$, which arises due to a single buyer switching to that choice, is large enough to cause (on average) $\alpha \ge 1$ buyers to switch to the same choice.  This causes an avalanche of decision changes which leads to a condensation of the market onto a single good (i.e. $Q\sim 1$).  {\color{black} (At large $J$, the market can (if pushed in the right direction) condense onto \emph{any} good -- this is why there are multiple solutions to \eqref{eq:Ueq}).}  We can estimate that \cite{leelucas} \begin{equation}
\alpha \sim J \left|\frac{\partial q}{\partial p_i} \right| \sim J \frac{M^{-1}}{\sigma (\log \frac{M}{2})^{-1/2}} = \frac{J}{\sigma} \frac{\sqrt{\log \frac{M}{2}}}{M} \equiv \bar J.
\end{equation}
Hence the PC phase exists for $\bar J \ll 1$, but not for $\bar J \gg 1$.

To deduce what happens when $\bar J \gg 1$, we must think about dynamics.  At early times, an emergent monopolist will capture the market.  As in Figure \ref{fig:mono}, if they capture the entire market, they will overprice their good (see SI) by not anticipating how much of their good's value derives from network effects, and thus precipitate a market crash, leading to their market share decaying as $(1-\rho)^t \approx \mathrm{e}^{-\rho t}$ (for $\rho \ll 1$).  However, if the buyers are sufficiently slow, the sellers will ramp up their price more slowly, which allows them to more accurately estimate the demand curve.  This monopolist will maintain market share  when $\bar \rho \ll 1$, where (see SI)
\begin{equation}
\bar \rho = \frac{\rho M}{\log M}.
\end{equation}
In contrast, when $\bar\rho \gg 1$, the market crashes due to overaggressive pricing.   Once the monopolist has lost market share, network effects will again drive an instability wherein a different seller will capture the market share and become an emergent monopolist.  This cycle of market condensation and crashes forms a \emph{non-equilibrium} (NE) phase, in which the permutation symmetry is broken at any fixed time, but is restored on long times, in the sense that each seller will have the same time averaged behavior: e.g. \begin{equation}
   \lim_{T\rightarrow\infty}\frac{1}{T}\sum_{t=1}^T q_i(t) = \frac{1}{M}.
\end{equation} A useful dynamical order parameter for the NE-SB phase transition is the rate $\gamma$ at which the seller with the largest market share changes, which is finite in NE and 0 in SB.  

We detail two finite size effects, quite visible in simulations, which can appear to modify the parameters where phase transitions occur.   Let $\bar J_{\mathrm{c}}$ denote the critical value at which the buyers would (in the absence of seller dynamics) condense into a single seller's good. In our model $\bar J_{\mathrm{c}}\approx 0.7$.   First, suppose that $M,N$ are very large, and $\bar J - \bar J_{\mathrm{c}} \ll \bar J_{\mathrm{c}}$. If \begin{equation}
\frac{1}{\bar J - \bar J_{\mathrm{c}}} \frac{1}{\bar \rho} \gtrsim \frac{\log M}{\log N}, \label{eq:finiteNstabilized}
\end{equation}
a seller raises prices fast enough during the ``decision change avalanche" described above that they significantly slow down the avalanche.  The market then mostly appears to be in PC, but exhibits extreme and short-lived bursts in single seller market share when (by random chance) a seller does not update prices for a long time. We associate such dynamics with the NE phase, but the transition from PC is continuous and can appear quite slow (see Figure \ref{fig:dynamic}).  In contrast, when $N$ is ``small" and $M>2$, finite-size fluctuations in prices $p_i$ can push the market out of the \emph{metastable} \cite{leelucas} equilibrium of PC; we find this occurs if (see Appendix \ref{app:outofPC})
 \begin{equation}
 \bar J_{\mathrm{c}} - \bar J \lesssim  M^{1/3}N^{-1/3}. \label{eq:Jfinitesize}
 \end{equation}
 Here the transition to NE or SB appears nearly discontinuous.

\begin{figure*}[t]
\includegraphics[width=0.6\textwidth]{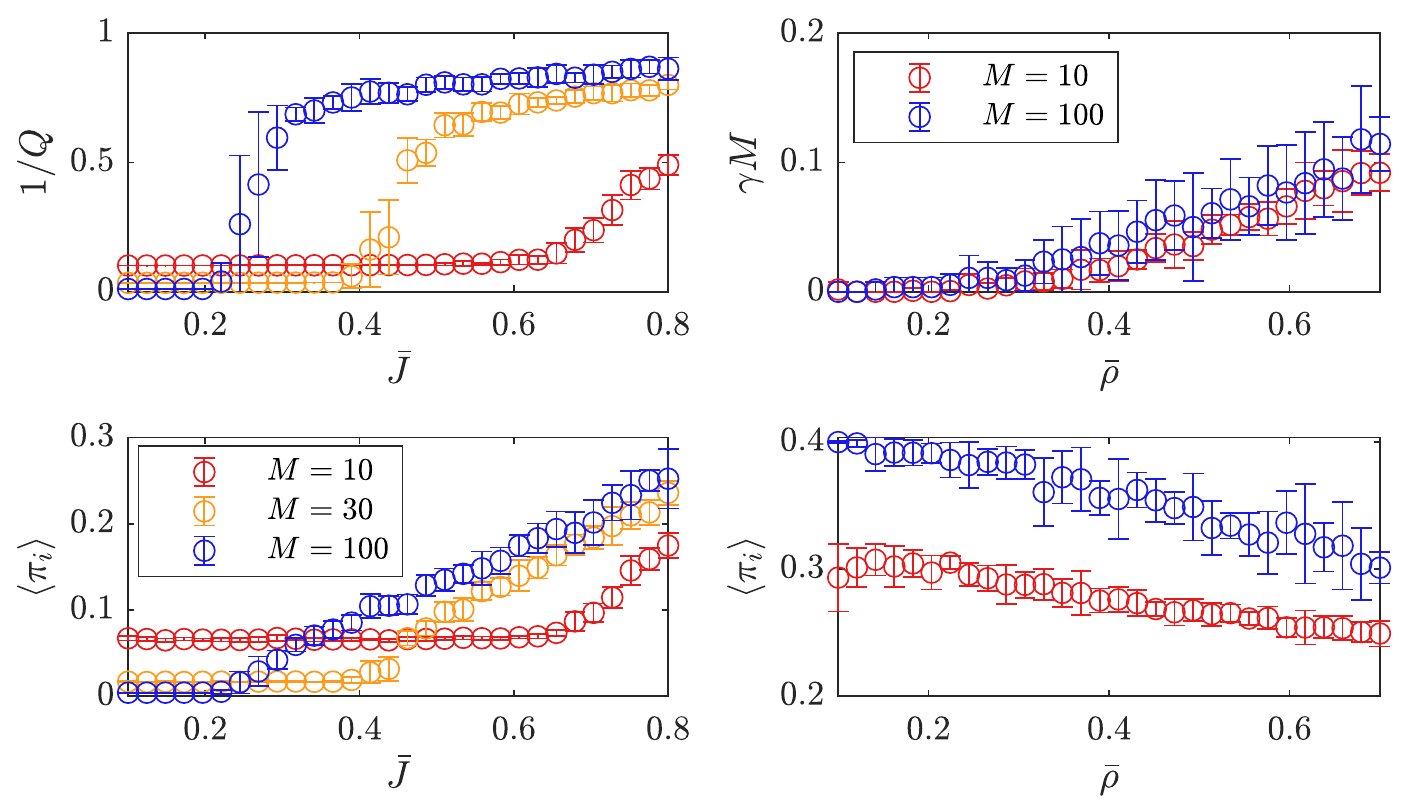}
\includegraphics[width=0.335\textwidth]{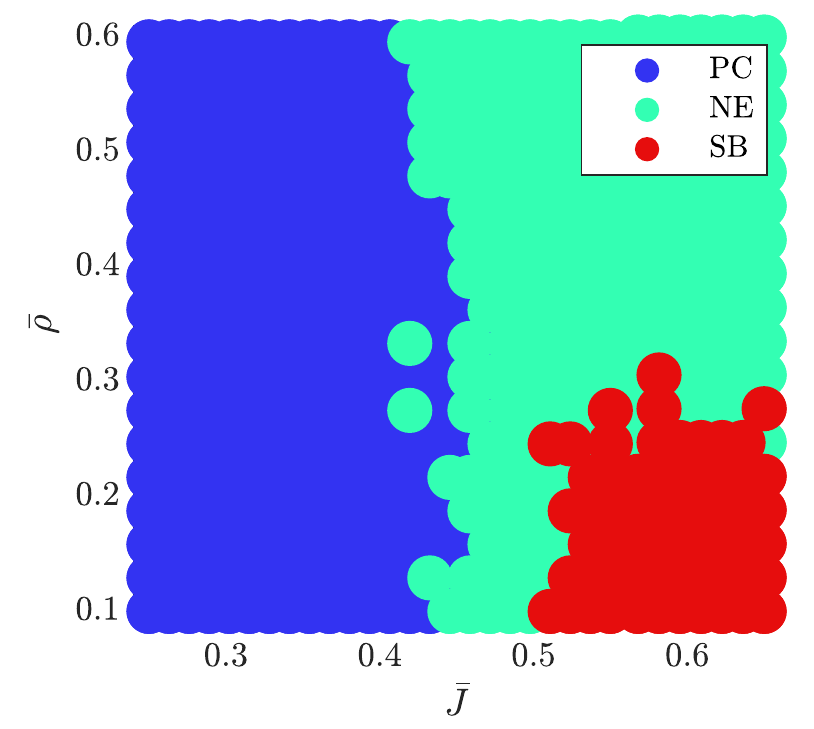}
\caption{\emph{Left:} The NE-PC transition at fixed $\bar\rho=0.8$, visible in both average seller profit $\langle \pi_i\rangle$ and in $M/Q$.   Corrections to the critical point due to finite $N/M$ are clearly visible.  \emph{Middle:} The NE-SB transition at fixed $\bar J = 0.9$, visible in $\gamma M$; we also show that the industry-averaged profit is finite in both phases.  \emph{Right:} Numerical determination of the phase diagram of the model in the $(\bar J,\bar\rho)$ plane at $M=30$.  To classify each point to a phase, we demanded $Q > M/3$ for the PC phase, and a median flip rate $\gamma =0$ for the SB phase between 9 random realizations.    All simulations used $N=100M$ and $N$ time steps.}
\label{fig:trans}
\end{figure*}

As a semantic point, therefore, any thermodynamic limit of $N,M\rightarrow \infty$ in which the phases described above are well-defined must be taken carefully, with the limit $N\sim M^a \rightarrow \infty$ taken together in a suitable fashion.  This technicality aside, agent-based simulations agree well with our expectations.   Figure \ref{fig:dynamic} shows the behavior of $q_i(t)$ and $p_i(t)$ in one model realization in various phases of the model. Sellers overpricing goods and causing market crashes is easily visible as the mechanism behind persistent dynamics in NE.  Figure \ref{fig:trans} demonstrates that the advertised order parameters behave as expected near the PC-NE transition as well as the NE-SB transition, along with a phase diagram of our model in the $(\bar J, \bar \rho)$ plane.  Finite size corrections to the critical $\bar J_{\mathrm{c}}$ where PC ceases to exist are significant, but are consistent in magnitude with basic estimates, as shown in Appendix \ref{app:outofPC}.

Appendix \ref{app:NESB} details the dynamical nature of the NE-SB transition, and discusses the microscopic origin of the transition.   Appendix \ref{app:ssb} argues that there will never be spontaneous symmetry breaking beyond that observed in the NE/SB phases (in the absence of production costs).  

{\color{black} If one incorporates production costs into the model, the qualitative phase diagram does not change.  In fact, since a standard assumption is that the production costs $w(q)$ -- which modify the firm's profit to $\pi_i = p_i q_i - w(q_i)$ --  obey $\mathrm{d}^2w/\mathrm{d}q^2 < 0$, we might expect that production costs \emph{destabilize} PC, because sellers who (by some random fluctutaion) get a few extra buyers can lower their cost relative to other sellers even further, while simultaneously the network effects also drive more buyers to this same seller.  This enhances the tendency of buyers to condense into purchasing from a single seller.  This effect is realized in our simulations: details and results from numerical simulations are presented in Appendix \ref{app:prodcost}.

A second extension to the model is to include noise in buyer/seller decisions.  There are three types of noise that we consider in Appendix \ref{app:noise}:  (\emph{1}) sellers can only sample a fraction of buyer preferences when setting their profit-maximizing price; (\emph{2}) sellers set their price at $p_i^*+\eta$, with $\eta$ a random variable; (\emph{3}) buyers do not always pick the highest utility choice.  (\emph{1}) and (\emph{2}) are found empirically to destabilize PC: in the former case because small $N$ effects, described by \eqref{eq:Jfinitesize}, are amplified in seller behavior, and in the latter case because for moderately large $\bar J$, PC is a metastable phase and the larger seller price swings may push the market out of the permutation-symmetric stability basin for the buyer dynamics.  In contrast, (\emph{3}) stabilizes PC, analogous to how finite temperature stabilizes a disordered permutation-symmetric phase in a random-field Potts model.}

\section{Distribution of Market Shares}

\begin{figure}[t]
\includegraphics[width=\columnwidth]{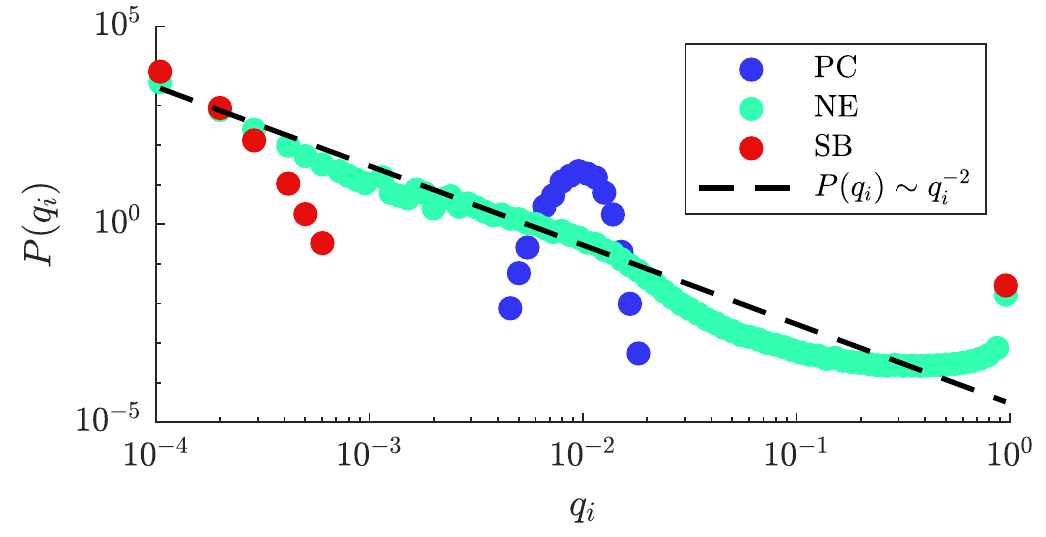}
\caption{$P(q_i)$ measured in 10 simulations with $N=10000$, $M=100$, $\mu=5$, $\sigma=1$.  PC:  $\bar J = 0$, $\bar \rho = 0.6$.   NE:  $\bar J = 0.9$, $\bar \rho = 0.6$. SB:  $\bar J = 0.9$, $\bar \rho = 0.05$.  We observe heavy-tailed $P(q_i)$ in NE; sharply concentrated $P(q_i)$ in PC; and a highly bimodal distribution near $q_i=0,1$ in SB.  Note that $P(0)$, not displayed, is substantial in SB for this $N,M$.}
\label{fig:qdist}
\end{figure}

Having established the phase diagram in our model, we now predict heavy-tailed distributions in the distribution of sellers' market shares, $q_i$, in the NE phase.  While part of this tail simply arises from the emergent monopolist, we predict heavy tails in the distribution \emph{even among less popular sellers}.  When a monopolist loses market share, the newly free buyers will select their next seller $i$ at a rate proportional to $u_{\alpha,i} + Jq_i - p_i$.    $p_i$s and $u_{\alpha,i}$ may be similar for all sellers, but the $Jq_i$ term suggests a preferential attachment (``rich get richer") mechanism, whereby a seller who just happens to have a large market share will gain an even larger one with time (and so fluctuations in $q_i$ get amplified with time).   Numerically, we confirm that the probability density $P(q_i)$ of market shares {\color{black} has heavy tails.  If one desires to fit to a power law, the best fit appears to be} roughly $P(q_i) \sim q_i^{-2}$. Curiously, this exponent is known to appear in the preferential attachment model of \cite{redner}, but its applicability to our model is unclear: in particular,  \emph{growing} systems obtain power-law distributions, yet the number of sellers in our model is fixed.   Regardless of the microscopic origin, in our simulations, {\color{black} heavy tails are present over a broad range of scales, and (see SI) should be expected over at least the range $M^{-2}\lesssim q_i \lesssim M^{-1}$.  The heavy tailed distribution in NE sharply contrasts with PC, where} $P(q_i)$ is concentrated around $q_i\approx M^{-1}$.   In SB (besides an obvious spike for $q_i \sim 1$) we find a more rapidly decaying tail at small $q_i$: see Figure \ref{fig:qdist}.  Numerically, the sharpest $P(q_i)\sim q_i^{-\nu}$ (with $\nu \approx 2 $) scaling occurs near the NE-SB transition (see Appendix \ref{app:pqi}).

It is empirically established \cite{mata, garicano} that there is a heavy tail in the market shares of firms; this is known as Gibrat's Law.  Economists commonly deduce this using e.g. firm employee counts; what is important is the broad distribution of firm sizes that arises in real markets.  Our model, which accounts for very little of the complications of supply-side economics, already includes one microscopic mechanism for these heavy-tailed distributions.  
%Combined with persistent oscillations in $q_i(t)$ and $p_i(t)$, heavy-tailed $P(q_i)$ together with persistend constitute the strongest test for our model in economic data.  

\section{Universality of NE phase}

Let us now argue that the \emph{qualitative conclusions} of our model can be robust; namely, even if seller pricing strategies are rather different, profit-maximizing sellers will generically drive the market out of the PC phase into either the NE or SB phases.  {\color{black} Firstly, as noted above, the qualitative nature of the phase diagram with PC at small $\bar J$ and NE or SB at large $\bar J$ are unchanged in the presence of noise or production costs.}  More broadly, in the spirit of effective theories in physics, consider each seller maximizing \begin{equation}
\Pi_i = \int\limits_0^T \mathrm{d}t \left(\pi_i(p_i;p_1,\ldots, p_N)- \zeta \left(\frac{\mathrm{d}p_i}{\mathrm{d}t}\right)^2+ \cdots \right),
\end{equation}
where the phenomenological parameter $\zeta$ captures stiffness or locality in price dynamics.  We can think of $\Pi_i$ as a ``(negative) action" (a la Lagrangian mechanics) which seller $i$ wishes to maximize.  Let us schematically carry out this maximization.  In the PC phase $q_i \sim M^{-1}$ and $p\sim \sigma (\log \frac{M}{2})^{-1/2}$ are constant in time, thus leading to  total profit $\Pi \sim \sigma M^{-1} (\log \frac{M}{2})^{-1/2} T$ in time $T$, if seller $i$ prices similarly.  However,  if the seller can drive the market to a symmetry broken point, then (if $\eta \rightarrow 0$) in the NE phase they can make profit $\Pi \gtrsim M^{-1}J T$ by setting price $J$ whenever they capture the market (on average for a time $T/M$).  Thus whenever $\bar J \gtrsim 1$, forward-thinking profit-maximizing sellers drive the market into the NE phase.   
Of course, a sufficiently wise seller may subsequently stabilize the market in a SB phase once they capture the market \cite{gaskins, cabral, grilo}: the emergent monopolist could set a lower price to avoid market crash. But if sellers need to learn the value of $J$, there may be a long period of oscillatory dynamics before a permanent monopolist arises.  
%Moreover, we numerically find a short time delay between a seller raising price and the avalanche of buyers switching to another good (at $\bar \rho \sim 1$), suggesting that sellers may struggle to avoid overshooting the optimal price in real time. 

\section{Conclusions}

There are many extensions of this model which will be important to analyze in order to further verify the robustness of our conclusions against features of a real economy we have not yet accounted for.  (\emph{1}) Supply-side constraints -- notably that sellers may need to produce their goods in advance of buyers choosing them (thus necessitating estimates of future demand) are not included yet. The model described here may best model markets for software downloads and/or online entertainment, where production costs or infrastructure are less critical and sellers can instantaneously provide a good to a buyer who pays for it.  Alternatively, the persistence of bursty fad-driven dynamics is consistent with empirical observations in fashion markets \cite{sproles}.  (\emph{2}) A game-theoretic analysis of optimal  seller strategies in this model, accounting for both network effects and the strategies of other sellers \cite{cabral}, is needed.  It is unclear if the NE phase could still exist as $t\rightarrow \infty$ without technological growth \cite{baumol}, or buyer/seller entry/exit into the market.  It is tempting to speculate that, analogously to repeated prisoner's dilemma games where cooperation can be desirable \cite{axelrod,sigmund}, in a competitive market with large network effects, firms may learn to accept oscillatory dynamics in exchange for being assured a larger time-averaged profit than in a competitive phase.   (\emph{3}) Heterogeneous non-equilibrium phases that arise due to locality  on a buyer/seller interaction graph are likely to exist, as in other evolutionary games \cite{nowak}. (\emph{4}) Including firm heterogeneity may lead to broad distributions in firm \emph{growth/decay rates}, as well as in $P(q_i)$ \cite{stanley}. (\emph{5}) Including macroeconomic market dynamics coupling the markets for many goods together \cite{delligatti, gualdi, jpb20} is important.    

Although our cartoon model is certainly incomplete as a model of an actual competitive market, the essential point of this paper is that network effects could generically lead to {\color{black} an unexpected failure of economic lore: that a market of memoryless utility-maximizing buyers and profit-maximizing sellers may \emph{never} reach any static equilibrium, exhibiting unpredictable dynamics for all time.  This conclusion holds even without any exogenous shocks.  We emphasize that this phenomenon is not equivalent to chaos that can occur when buyers maximize a utility function $u(x(t-1),x(t))$ that depends on choices at two (or more) times \cite{boldrin}: in our model, buyers only gain utility from their current decision and, without feedback from sellers, reach equilibrium even with network effects. The long run analysis of even highly competitive markets will be misleading if equilibrium is unstable.}  If the macroeconomy is already fragile even in the absence of network effects \cite{acemoglu, jpb20}, it is all the more crucial to understand whether network effects play an important role in persistent price dynamics observed in real economies.  

We found that the NE phase will exhibit both the strongest temporal fluctuations in prices and the broadest distributions of firm's market shares; looking for the presence or absence of this correlation in empirical data may be a simple test of our model.  Following \cite{summers,sornette,Marcaccioli_2022}, a more quantitative check for whether fluctuations are endogenous (intrinsic to the market) or exogenous (driven by external shocks) is to study the time correlations of market properties such as $Q(t)$. In particular, in our model, exogenous shocks in the PC phase cause rapid drops in $Q(t)$ followed by slow increases as the market returns to equilibrium.  In contrast, in and near the NE phase, the largest jumps in $Q(t)$ are when it \emph{increases} abruptly as the dominant seller abruptly loses market share.  In Appendix \ref{app:shock}, we confirm this expectation quantitatively in numerical simulations using time correlations in $Q(t)$. Hence, with neither the ability to control network effects, nor knowledge of the times or strengths of exogenous shocks, it may be possible to confirm our theory for a competitive market, driven far from equilibrium by endogenous fluctuations, by studying dynamics of prices and market shares in real world markets.

\section*{Acknowledgements}

I thank Jean-Phillippe Bouchaud for useful comments on an early draft. This work was supported by the Alfred P. Sloan Foundation through Grant FG-2020-13795.

\begin{appendix}
\section{The PC phase}\label{app:PC}
In this appendix we derive some of the details about the perfect competition phase claimed in the main text, along with others that are useful for remaining appendices.

Let us first argue why the typical seller price \begin{equation}
p_i \sim \frac{\sigma}{\sqrt{\log M}}. \label{eq:psqrtlog}
\end{equation}
A typical seller $i$ will aim to set a price $p_i>0$ such that even if the price of all other sellers were exactly zero, they would still keep some buyers:  $q_i>0$ and thus $\pi_i>0$.  The price at which they can achieve this criterion is such that given the $M$ Gaussian random variables $u_i$, ordered such that $u_1>u_2>\cdots >u_M$, the difference $p_i \sim u_1 - u_2$.   To estimate $u_1-u_2=z$, let us take $\mu=0$ and $\sigma=1$ for simplicity.   Using extreme value statistics, the probability density function $\rho(z)$ is given by \begin{equation}
\rho(z) = M(M-1) \int\limits_{-\infty}^\infty \mathrm{d}u F(u)^{M-2}F^\prime(u)F^\prime(u+z).
\end{equation}
Here $F(u)$ is the cumulative distribution function for a Gaussian random variable.   We now estimate the scaling of $\rho(z)$ via saddle point.  By extreme value theory we know that the integral will be dominated for $u\sim \sqrt{\log M}$:
\begin{align}
 &\int\limits_{-\infty}^\infty \mathrm{d}u F(u)^{M-2}F^\prime(u)F^\prime(u+z)  \notag \\
 &\sim  \int\limits_{-\infty}^\infty \mathrm{d}u  \exp\left[ -M \frac{\mathrm{e}^{-u^2/2}}{u} - \frac{u^2 + (u+z)^2}{2}\right].
 \end{align}
 The saddle point equation is \begin{equation}
M\mathrm{e}^{-u^2/2} = 2u+z,
 \end{equation}
 whose solution is (at leading order and when $z\rightarrow 0$): \begin{equation}
 u \approx \sqrt{2\log \frac{M}{2} - \log \log \frac{M}{2}} \equiv \lambda.
 \end{equation}
We then estimate that\begin{equation}
\rho(z) \approx \lambda \mathrm{e}^{-\lambda z}. \label{eq:rhoz}
\end{equation}
 or that the typical value of $z \sim (\log M)^{-1/2}$.  
 
The scaling of $\lambda$ is sensitive to the microscopic distribution on $u_{\alpha,i}$.  For example, suppose we instead have \begin{equation}
\mathbb{P}(u_{\alpha,i} > x) = \mathrm{e}^{-x}; \label{eq:exprv}
\end{equation}
a simple calculation reveals that \emph{exactly}
 \begin{equation}
\rho(z) = \mathrm{e}^{-z},
\end{equation}
since the conditional form of the cumulative distribution function for these exponential random variables is identical to a shifted (\ref{eq:exprv}).  In a market with this distribution function, a seller would set their prices at $p_i \sim 1$, independently of $M$.  If the utility distribution is heavy-tailed then even without network effects, we can expect that seller profit would \emph{grow} with market share, and this could be interesting to investigate further.

\begin{figure}
 \includegraphics[width=\columnwidth]{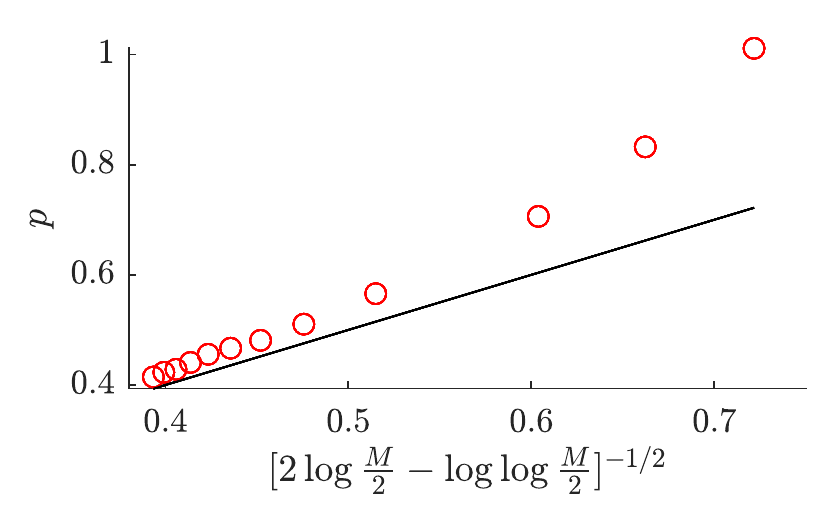}
 \caption{The time- and seller-averaged price $\langle p_i\rangle$ vs. $M$ in the PC phase with $J=0$, and $N=100M$.  Red denotes perfect scaling with (\ref{eq:psqrtlog}) and black is data from simulations.  We have set $\sigma=1$, and averaged over 40 simulations for each data point.}
 \label{fig:pricesPC}
 \end{figure}

If $N\rightarrow \infty$ (and sufficiently large $M$ so that our extreme value statistics is accurate), sellers will fix their prices to be \begin{equation}
p_i = \lambda^{-1}. \label{eq:pilambda}
\end{equation}
To understand why, as in textbook economics, note that all sellers will choose exactly the same price, since they are statistically equivalent and there are no demand-side fluctuations as $N\rightarrow \infty$; if any one seller could lower price and gain more market share (and in the process make more profit) they would do so.    Therefore, each seller must set their price such that when only the buyers whose favorite good is theirs do in fact select their good, they are also maximizing profit.   Thus, we need \begin{equation}
\frac{\mathrm{d}\pi}{\mathrm{d}p} = 0 = \rho(0) + p\rho^\prime(0) = 0,
\end{equation}
which reduces to (\ref{eq:pilambda}).  Figure \ref{fig:pricesPC} confirms that the average price in the PC phase in large-scale simulations largely follows our predictions when $M$ is large.

\begin{figure*}
 \includegraphics[width=\textwidth]{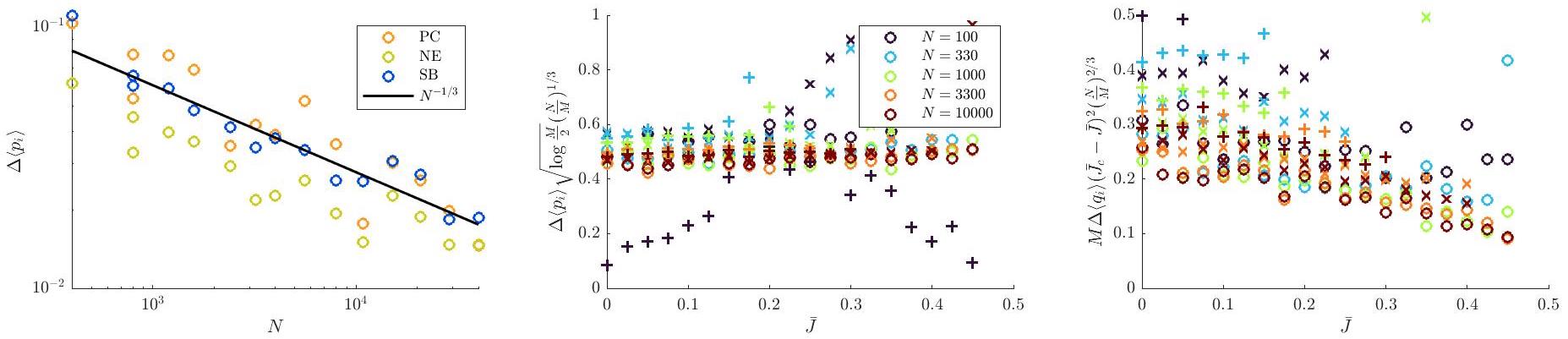}
 \caption{\emph{Left:} Simulations with $M=4$; which is the \emph{sample-to-sample average} of the average seller prices, averaged over time.  In all phases we see $\mathrm{\Delta}p\sim N^{-1/3}$.  Parameters used are: PC $(\bar J=0, \bar \rho = 0.5$), NE ($\bar J = 3, \bar \rho = 0.5$), SB ($\bar J=3, \bar \rho = 0.1$), with $\mu=4\sigma$.  \emph{Middle/Right:} Simulations of $\mathrm{\Delta}p_i$ and $\mathrm{\Delta}q_i$ confirming the qualitative predictions (\ref{eq:deltap}) and (\ref{eq:deltaq}).  Circles denote $N=10$, Xs denote $M=30$, and +s denote $M=100$.  While $\mathrm{\Delta}p_i$ appears to closely track our prediction (up to a constant offset) once $N\gtrsim1000$ (the strongest deviations occur when $N$ and $M$ are comparable), we see notable deviations from our $\mathrm{\Delta}q_i$ prediction.   We expect this is a consequence of seller behavior which suppresses fluctuations in $q_i$ when $N/M$ is very large.}
 \label{fig:N}
 \end{figure*}

 Lastly, let us discuss the fluctuations in prices in the PC phase at finite $N$.  As viewed by a single seller, they will see a ``noisy" demand curve which we can heuristically model as \begin{equation}
q(p) = q_0(p)+ \sqrt{\frac{1}{MN}}\int\limits_{0}^p \mathrm{d}p^\prime \sqrt{q_0^\prime(p)} \xi(p^\prime),
 \end{equation}
 where $q_0(p)$ corresponds to the $N=\infty$ demand curve (essentially $\rho(z-\langle p_i\rangle)$), and $\xi$ corresponds to Gaussian white noise:  $\langle \xi(p)\xi(0)\rangle = \delta(p)$.     The scaling in this white noise can be deduced as follows:  integrating the noise term from $p=0$ to $p=\langle p\rangle$, fluctuations in $q(p)$ in the PC phase must be of order $1/\sqrt{MN}$, since all sellers in the thermodynamic limit would (with all prices identical) simply pick the seller with the highest $u_{\alpha, i}$.  This leads to a binomial distribution on $Nq(p)$ with mean $N/M$ and variance $\sqrt{N/M}$.  We deduce the seller's profit maximizing price as follows:  suppose that at finite $N$, a seller deviates from the market average by an amount $\mathrm{\Delta} p$; then we can estimate their profit as \begin{equation}
 \pi(p) \sim  \frac{\langle p\rangle}{M} -   \frac{(\lambda \mathrm{\Delta}p)^2}{M} + \sqrt{\frac{\lambda \mathrm{\Delta}p}{MN}}.
 \end{equation}
 where we have neglected O(1) constants in the above estimation.  The second term above is estimated by noting that when $\mathrm{\Delta}p \sim \lambda^{-1}$, the seller will lose a finite fraction of their $1/M$ market share.  The third term above comes from estimating the stochastic constribution to $q(p)$.  We deduce optimizing the formula above that \begin{equation}
\lambda  \mathrm{\Delta}p \sim \sqrt{\frac{M}{N}},
 \end{equation}
 or using (\ref{eq:pilambda}), \begin{equation}
 \mathrm{\Delta}p \sim \frac{\sigma}{\sqrt{\log \frac{M}{2}}} \left(\frac{M}{N}\right)^{1/3}. \label{eq:deltap}
 \end{equation}
 We confirm this scaling in numerics in Figure \ref{fig:N}.  Amusingly, it seems to hold reasonably well in the NE and SB phases too.
 
 This change in prices leads to much stronger fluctuations in $\mathrm{\Delta}q_i$ at finite $N$ than one would naively estimate from the ``demand curves" alone.  Indeed, we can approximate that \begin{equation}
 \mathrm{\Delta}q \sim \frac{\lambda \mathrm{\Delta}p}{M},
 \end{equation}
 or that (using $Q$ defined in the main text) \begin{equation}
 \frac{M}{Q}-1 \sim (M\mathrm{\Delta}q)^2 \sim \left(\frac{M}{N}\right)^{2/3}. \label{eq:deltaq}
 \end{equation}
 This scaling is also confirmed in Figure \ref{fig:N}.

 \section{The PC-NE/SB transition at small and large $N$}\label{app:outofPC}
 Now we turn to a discussion of the demand side of the instability when $\bar J >0$.  First, let us estimate the critical value $\bar J_{\mathrm{c}}$ above which the buyers will drive the market into a symmetry-broken phase.   This is done by demanding that \begin{equation}
 1 = \frac{M}{M-1}  J_{\mathrm{c}} \left|\frac{\partial q_i}{\partial p_i}\right|, \label{eq:Jcdef}
 \end{equation}which says that on average each buyer that switches seller will induce 1 other buyer to switch seller (leading to an infinite expectation value for the number of sellers that switch and heralding the $N=\infty$ location of the phase transition:  see \cite{leelucas} for extensive further discussions/interpretations).   The factor of $M/(M-1)$ comes from the fact that the good the buyer switched from itself becomes \emph{less popular}.
 
 There are two limits where we can do this calculation analytically.  If $M=2$, then we know that $q_i(p_i)$ is normally distributed with standard deviation $\sqrt{2}\sigma$ (as this is the distribution of $u_{\alpha,1}-u_{\alpha,2}$).    Using (\ref{eq:Jcdef}) and noting that $\partial q/\partial p$ is nothing but the probability distribution function of the Gaussian, we conclude that \begin{equation}
 1 = 2  J_{\mathrm{c}} \times \frac{1}{\sqrt{2\pi \cdot 2}\sigma }
 \end{equation}
 or that \begin{equation}
 J_{\mathrm{c}} = \sqrt{\pi} \approx 1.77\sigma .
 \end{equation}
 Alternatively, when $M\rightarrow \infty$, we can use the calculation of (\ref{eq:rhoz}) to deduce that $\partial q/\partial p = \lambda$.  Comparing to the definition of $\bar J$, we conclude that \begin{equation}
 \bar J_{\mathrm{c}} = \frac{1}{\sqrt{2}} \approx 0.707.
 \end{equation}
 We will use these estimates when making plots, for simplicity, although we do caution that the latter has finite $M$ corrections.
 
\begin{figure}
 \includegraphics[width=\columnwidth]{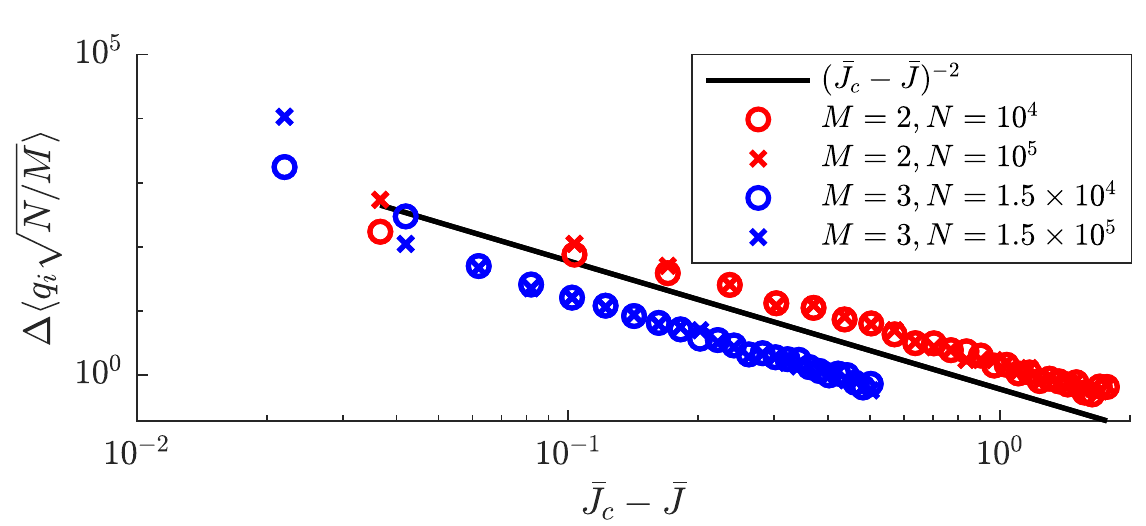}
 \caption{Confirming the scaling (\ref{eq:q1alpha}) in demand-only simulations (i.e. all prices $p_i$ are equal).  The offset is due to the change in $J_{\mathrm{c}}$ between different $M$.  Note that we numerically find $\bar J_{\mathrm{c}}\approx 0.502$ for $M=3$.}
 \label{fig:alpha}
 \end{figure}

 Equipped with the knowledge of $\bar J_{\mathrm{c}}$, let us now consider the state of the market with a small number $N$ of buyers.  There will be intrinsic fluctuations in the initial conditions \cite{leelucas} due to finite size effects;  let us write \begin{equation}
 \delta q_i = q_i - \frac{1}{M}
 \end{equation}
 to denote these small effects.  In the presence of network effects, we expect that for $\bar J < \bar J_{\mathrm{c}}$, 
  \cite{leelucas} \begin{equation}
  \delta q_i(\bar J) = \frac{\bar J_{\mathrm{c}}}{\bar J_{\mathrm{c}}-\bar J}  \delta q_i(\bar J=0). \label{eq:q1alpha}
  \end{equation}
 Figure \ref{fig:N} shows that this scaling approximately holds in actual simulations of a market with dynamical sellers, albeit with somewhat notable corrections at large $N/M$.  We believe the reason for this is that at large $N/M$, the sellers dynamically suppress fluctuations (as we will explain below in more detail).   Figure \ref{fig:alpha} shows that this scaling indeed occurs when all sellers sell at the same price (which is low enough that essentially all buyers remain in the market) and when we allow buyers to ``relax" to equilibrium.
 
 These finite $N$ effects can be quite important even for the relatively large ratio $N/M=100$ that we commonly employed in our numerics.   A particularly drastic consequence of the fluctuations (\ref{eq:q1alpha}) is that they tend to (noticably!) decrease the value of $\bar J_{\mathrm{c}}$ well below its theoretically predicted value.   To understand why, note that once $M>2$, the demand-side phase transition from the symmetric point to a point where one seller spontaneously captures much of the market is discontinuous.  Therefore, even for $\bar J<\bar J_{\mathrm{c}}$, there are stable solutions which break symmetry.   
 
 We can roughly estimate the ``size" of the attractive basin of the symmetric point as follows, assuming $\bar J \ll \bar J_{\mathrm{c}}$ for simplicity.   Suppose that the market might condense onto seller $j$ (i.e. so $q_j \sim 1$).   This process will occur (assuming that all sellers keep their prices fixed) once the typical number of buyers induced to switch to $j$ (call it $\alpha$) by another buyer switching to $j$ obeys $\alpha =1$.   Using (\ref{eq:Jcdef}) (and taking $M$ large for simplicity) we conclude that $j$ can capture the market once \begin{equation}
 \frac{1}{J} = -\frac{\partial q_j}{\partial p_j} \sim \frac{\lambda}{M} \mathrm{e}^{\lambda J (q_j - 1/M)}
 \end{equation}where the last step above comes from a similar estimate to (\ref{eq:rhoz}) for the number of buyers who would switch to $j$ if the added utility of the choice was $J(q_j-1/M)$.  We can re-write this equation as \begin{equation}
 \frac{1}{\sqrt{2}\bar J} \sim \mathrm{e}^{\bar J M \mathrm{\Delta}q_j},
 \end{equation}
 which tells us both that (as claimed before) $\bar J_{\mathrm{c}}\rightarrow 1/\sqrt{2}$ at large $M$, and also that the value of $\bar J$ at which the instability will arise is given by, near $\bar J_{\mathrm{c}}$, $M\mathrm{\Delta}q_j \sim1$.  Combining (\ref{eq:q1alpha}) and (\ref{eq:deltaq}) we conclude that there is a finite size effect driven demand-side instability to PC when \begin{equation}
     \bar J_{\mathrm{c}}-\bar J \sim \left(\frac{M}{N}\right)^{1/3}.
 \end{equation}
 We could not confirm this exact scaling numerically because (as we discuss momentarily) there is a competing  dynamical effect as $N/M \rightarrow \infty$ which suppresses the instability, but it is clear from e.g. Figure 3 of the main text that for fixed $N$, increasing $M$ substantially decreases the apparent value of $\bar J_{\mathrm{c}}$.
 
\begin{figure}
 \includegraphics[width=\columnwidth]{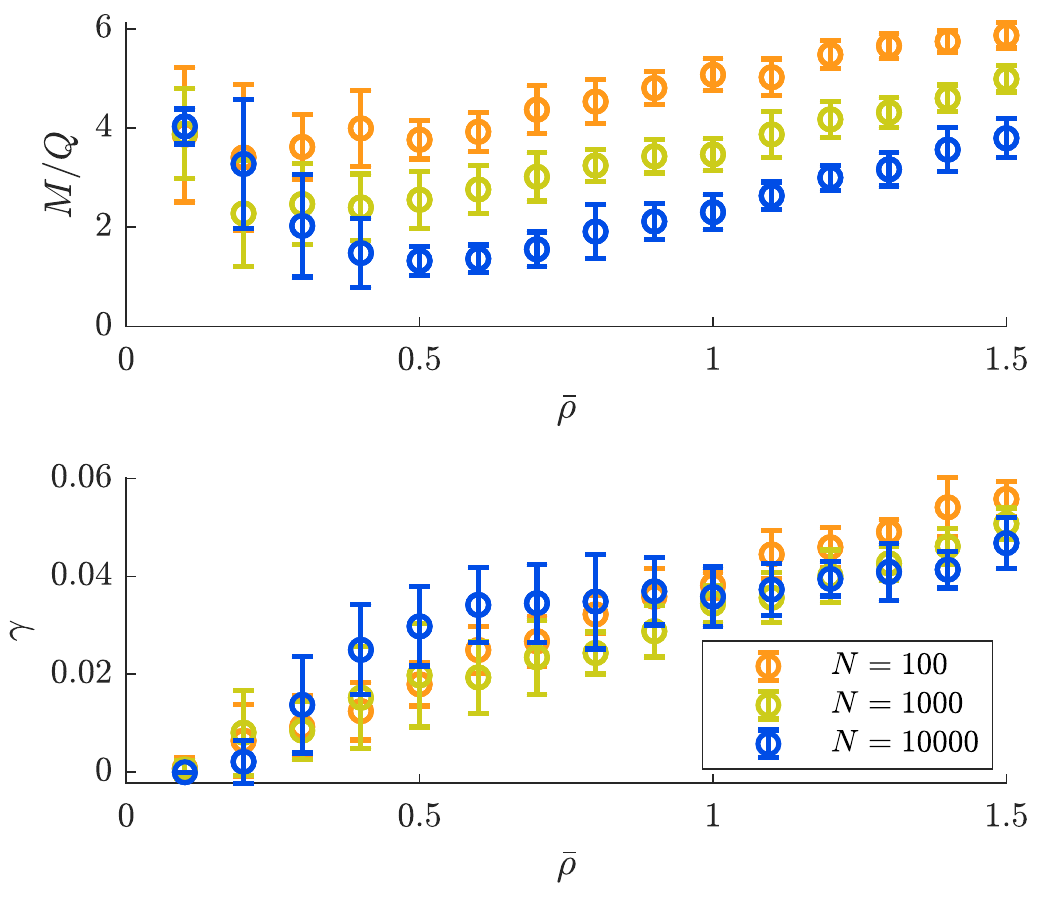}
 \caption{Simulations with $M=10$ sellers at $\bar J = 0.7$, very close to $\bar J_{\mathrm{c}}$.  For small $\bar \rho$ the model is in SB, and for large $\bar \rho$ in NE; however near the transition we find $Q$ is largest and gets increasingly large with increasing $N$.  This behavior is consistent with our prediction that seller dynamics can slow down market condensation in NE.}
 \label{fig:suppress}
 \end{figure}

 Now we describe how sellers can dynamically help to stabilize the symmetric equilibrium at large $N$.   Suppose that there were no seller price updates at all; how long would it take for the market to condense onto a seller for $\bar J > \bar J_{\mathrm{c}}$?  Very close to the transition, each seller induces of order $\bar J/\bar J_{\mathrm{c}}$ additional buyers to switch in the next time step, so we can estimate that (if seller $j$ captures the market) \begin{equation}
 \frac{\mathrm{d}q_j}{\mathrm{d}t} \sim  \rho \left(\frac{\bar J}{\bar J_{\mathrm{c}}}-1\right) q_j,
 \end{equation}
 suggesting it would take a time \begin{equation}
 t_{\mathrm{condense}} \sim \frac{\log N }{\rho (\bar J - \bar J_{\mathrm{c}})} \sim M  \frac{\log N }{\bar \rho \log M  (\bar J - \bar J_{\mathrm{c}})} 
 \end{equation}
 But sellers update their prices in a time $t\sim M$, and if this time scale is short compared to $t_{\mathrm{condense}}$, then seller price updates will significantly slow (or may even prevent) the market from condensing.   This leads to the criterion in the main text for when the critical point can be suppressed above $\bar J_{\mathrm{c}}$ at large enough $N$.  Some evidence for this effect, and perhaps even the $N$-scaling, is presented in Figure \ref{fig:suppress}, where we observe that upon increasing $N$ by powers of 10, there is a window at increasingly large $\bar J$ where the dynamics has an anomalously large value of $Q$ for $\bar J$ near $\bar J_{\mathrm{c}}$ and intermediate values of $\bar \rho$.
 
\section{The NE-SB transition}\label{app:NESB}
In this appendix we justify that the NE-SB transition is dynamical, and that if buyer dynamics is too fast the SB phase cannot exist in our model.

Let us begin by considering the behavior of a single seller who has captured the market.  If $\sigma=0$, as stated in the main text, the seller will fix price at $p=J-0^+$ and thus capture the market with profit $\pi=J-0^+$.  What happens when $\sigma$ is very small?  As the sellers in our model do not know to account for network effects, we can estimate\footnote{In reality, the distribution $q_0$ is not Gaussian, but must be conditioned on the price at which a seller would switch given their optimum seller $j$ comes with a utility $u_{\alpha,j}$ which is the maximum over $M$ Gaussian random variables.  This is a technical point that does not really modify the essentials of the argument, so we will go ahead and use the simpler model below.}  will choose their price by optimizing \begin{equation}
\pi(x) = (J+x)q_0\left(\frac{x}{\sigma}\right)
\end{equation}
where $x=p-J-\sigma\sqrt{2\log M}$ (the latter subtraction accounts for the fact that at large $M$, the typical buyer's top choice is another seller) and $-q_0^\prime(x) \sim \mathrm{e}^{-x^2/2}/\sqrt{2\pi}$ gives the demand curve which comes from the probability density function of the $u_{\alpha,i} - \mu$.  Looking for the maximum of $\pi$, and noting that $x \ll J$ when $\sigma$ is small:
 \begin{equation}
-\frac{J+x}{\sigma} q_0^\prime  \approx -\frac{J}{\sigma} q_0^\prime  = q_0 \approx 1. \label{eq:q0primeapp}
\end{equation}
We thus estimate that for $\sigma \ll J$, \begin{equation}
x \approx -\sigma \sqrt{2 \log\frac{J}{\sqrt{2\pi}\sigma}}. \label{eq:appBp}
\end{equation}
We conclude using the asymptotics of the error function that \begin{equation}
q_0 \approx 1 - \frac{\sigma}{J}\sqrt{\frac{2}{\log\frac{J}{\sqrt{2\pi}\sigma} }} \sim 1-\frac{1}{M\bar J}.
\end{equation}
But this loss in demand, due to network effects, makes buyers less willing to pay price $J+x$.  Also note that in this SB phase, the market share of the non-monopolist sellers will not vanish, but will instead be $\sim M^{-2}$; we confirmed this in numerics for relatively small $M$.

To estimate the consequences of this, we can think of how buyers would respond to a fraction $\delta$ of sellers leaving the market.  Based on that information, define the fraction of additional buyers that leave the market to be $\alpha\delta$.  If $\alpha \ge 1$, then the seller will lose all buyers at sufficiently late time.   We can now calculate $\alpha$ as follows: \begin{equation}
\alpha \delta = -q_0^\prime \left(\frac{x_0}{\sigma}\right) \frac{J}{\sigma} \delta.
\end{equation}We thus estimate there will be an instability since -- even if the buyers who stopped buying from the monopolist simply left the market,
\begin{align}
\alpha - 1  &\approx \frac{J}{J+x} - 1 \approx \frac{\sigma}{J} \sqrt{2 \log\frac{J}{\sqrt{2\pi}\sigma}} \notag \\
&\sim \frac{\sqrt{\log M}}{M\bar J} \sqrt{\log \frac{M\bar J}{\sqrt{\log M}}} \approx \frac{\log M}{M\bar J}.
\end{align}  
For simplicity we assumed $M\gg 1$ to estimate the asymptotics above.  Importantly, our estimate is that while $\alpha-1$ is small, it is larger than 1. This will inevitably tend towards instability if the monopolist does not act.  

  \begin{figure*}[t]
   \includegraphics[width=\textwidth]{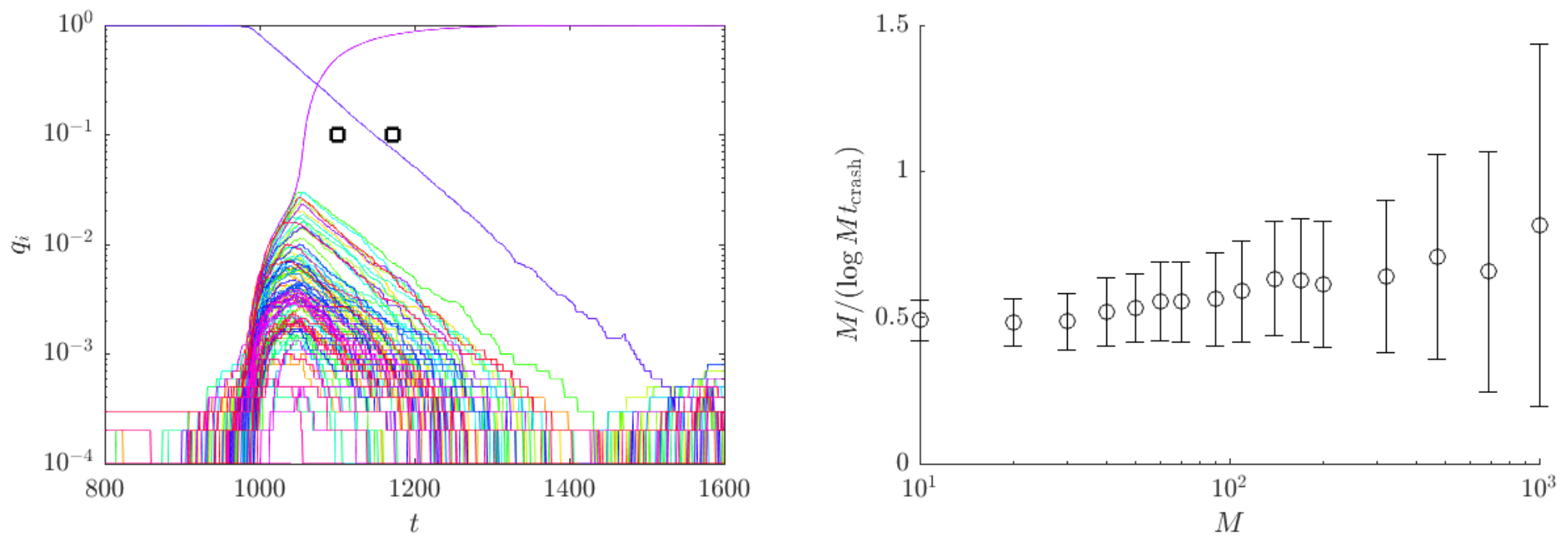}
 \caption{\emph{Left:} A model run at $M=100$ and $N=10000$, with $\mu=5$, $\sigma=1$, $\bar J=1$ and $\bar \rho =0.3$.   Different sellers are marked with different colors. We can observe clearly the exponential decay of $q_i(t)$ for sellers who are losing market share.  A broad and heterogeneous distribution of $q_i$ is also visible as the dominant seller in the market flips.  The two black squares denote the time interval $\rho^{-1}$, which is on the order of the time scale of the transition between two dominant sellers.  The dominant seller switches even when both have over 10\% of buyers, confirming that if the dynamical stabilization of the SB phase does not occur at late times, but instead at early times.  \emph{Right:} Numerical confirmation of the scaling $t_{\mathrm{crash}} \sim \rho^{-1} = M^{-1}\log M$ in numerical simulations with $N=10M$, $\bar\rho = 1$, $\bar J = 1$.  50-200 simulations were used for each data point.  In each simulation the market was initialized with all buyers in $i=1$, with seller $i=1$ fixing price only during time step $t=1$.}
 \label{fig:loglogqt}
  \end{figure*}

In fact, in Figure \ref{fig:loglogqt}, we show that if the monopolist has captured the whole market, they will overestimate their price in such a way as to precipitate a market crash in a time which numerically seems to scale close to $1/\rho$ -- namely, once each buyer has on average gotten to re-choose seller once, the seller has precipitated an inevitable market crash.  To understand why this is so fast relative to the very weak value of $\alpha$ predicted above, observe that the seller will want to set price such that $1-q_0 \sim (M\bar J)^{-1}$.  The fraction of buyers who flip in a given time window is (assuming $\rho  \ll 1$) \begin{equation}
\frac{\mathrm{d}q_0}{\mathrm{d}t} \approx -\rho \left(\frac{1}{M\bar J} + \alpha (1-q_0)\right),
\end{equation}
and since $\alpha$ is so small we can approximate it as close to 1.  This differential equation is solved by \begin{equation}
1-q_0 \sim \frac{\mathrm{e}^{\rho t} - 1}{M\bar J}.
\end{equation}
From (\ref{eq:appBp}), when \begin{equation}
1-q_0 \sim \frac{\log M}{M\bar J},
\end{equation}
then the monopolist's price is too high for the typical buyer and the market will crash.  This happens in a time \begin{equation}
t_{\mathrm{crash}} \sim \frac{\log \log M}{\rho},
\end{equation}
which we were unable to distinguish from $1/\rho$ in numerics.   We might naively conclude that when $t_{\mathrm{crash}} \lesssim M$, we will enter the SB phase.

However, as visible in Figure 3, it appears that a better proxy for the SB phase is $\bar \rho$, which has a factor of $\log M$ rather than $\log \log M$.   To understand why this is the case, we must think not about late time behavior, but instead about early time behavior.  Right as the monopolist is \emph{gaining} market share, we will find \begin{equation}
q_0(t) \sim 1 - \mathrm{e}^{-\rho t},
\end{equation}
where $t$ denotes the time after the seller begins to gain substantial share.   If the seller had market share $1-q_0\sim M^{-1}$ when setting their next price, then the extremely small value of $\alpha$ above would make it essentially impossible for the buyers' decision to cascade into a market crash before the seller could react and correct pricing.  Therefore, the dynamics will lead to an NE phase only when the buyer dynamics is so fast as to push $1-q_0 \ll M^{-1}$ before the seller can choose a different price.  This occurs when $\bar \rho \gg 1$, which justifies the scaling found in the main text.

  \section{Absence of further symmetry breaking}\label{app:ssb}
  In the models described thus far, one either finds an approximate $\mathrm{S}_M$ symmetry among sellers in the PC phase, or (at any fixed time $t$) an approximate $\mathrm{S}_{M-1}$ symmetry in the NE or SB phase.  (The symmetry is $\mathrm{S}_{M-1}$ since the market is essentially described by a monopolist with O(1) market share with $M-1$ sellers competing for very small market share.   In typical simulations we find that these sellers do not have $q_i=0$: see e.g. the dynamics in the SB phase in Figure 2.  
  
  Let us now argue that it is unlikely to ever see a further symmetry breaking pattern in the SB phase, at least without non-trivial distributions on the intrinsic utilities $u_{\alpha,i}$ \cite{leelucas}, or production costs (see Section 6 below).   Suppose there were a SB phase where 2 sellers $i=1,2$ had each captured O(1) market share.    Then from either seller $i=1,2$ perspective, they would want to maximize profit by making sure that (as in Appendix \ref{app:NESB}) $p\approx Jq_i$.   But e.g. seller 1 would be able to make profit $\pi_1 \approx (Jq_1 - \delta) (q_1+q_2) > Jq_1^2$ by lowering their price by $\delta \gtrsim \sigma$ to capture all of seller 2's market share.  This argument holds whenever $J \gtrsim \sigma$. Of course, we know that the PC phase is only unstable when $\bar J \gtrsim 1$, which always implies $J\gtrsim \sigma$; hence within our model we will never find heterogeneous distributions in firm size $q_i$ at a typical time.  It would be interesting to find generalizations of the model where this is possible.

 \section{Production Costs}\label{app:prodcost}
In this appendix, we describe the addition of explicit production costs to our model: namely, \begin{equation}
    \pi_i(q_i) = p_iq_i - w(q_i),
\end{equation}
where $w(q_i)$ denotes the production cost of selling to a fraction $q_i$ of all possible buyers.  One may interpret this model as asserting that buyers place an order for a good, such that the sellers do not worry about over-producing.  We expect that generalizing our model to one where sellers must predict their market share before the next time step does not completely destroy the phase diagram.

In economics it is commonly assumed that $w(q)$ (which we take to be the same for all firms) obeys $\mathrm{d}w/\mathrm{d}q>0$ and $\mathrm{d}^2w/\mathrm{d}q^2 < 0$ -- namely, for most goods it will always cost at least \emph{some amount more} to make the good, while a firm would never find the cost per good to increase with $q$ -- if so, they would simply cut their production into smaller ``units" to ensure the lower production cost per good.  A simple function which we find useful for our simulations is \begin{equation}
    w(q) = b \min \left(\frac{a}{M},q\right) + cq.
\end{equation}
This function says there is a constant production cost per good of $b+c$, until the seller sells to a fraction $a/M$ of the market -- after this, further production costs per good are only $c$.  This is, obviously, a very oversimplified function.  The reason why we choose it is that it is simple and it mirrors the desired generic properties of $w(q)$ discussed above (in a piecewise fashion).   Note that in our model, so long as buyers are not on the verge of dropping out of the market altogether, the constant $c$ will effectively just lead to prices $p\rightarrow p+c$: after all, the profit maximization problem at price $p$ finite $c$ is trivially equivalent to the same problem with $c=0$ but with seller price $p-c$.   As this does not change the phase diagram, in what follows, we therefore set $c=0$.

\begin{figure}[t]
    \centering
    \includegraphics[width=\columnwidth]{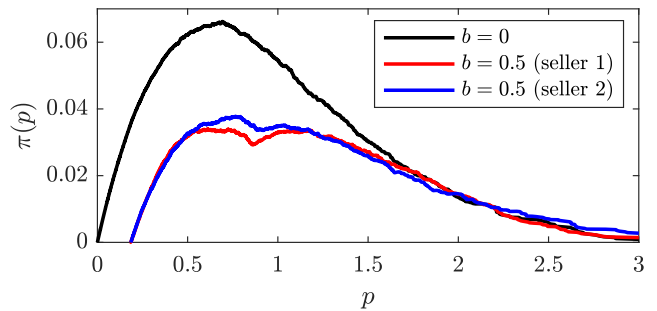}
    \caption{Numerical samples of the seller's optimization problem: plots of $\pi(p)$ as a function of $p$ for sample sellers at different $b$.  Simulations had $\bar J=0$, $M=10$, $N=10^4$, $\sigma=1$, $\mu=50$, $a=0.8$.  The seller can make profit for $p<b$ because at this price, the market share will be (significantly) larger than $a/M$.}
    \label{fig:prodprofits}
\end{figure}

We expect that including production costs will always enhance the instability of the PC phase.  To understand why, observe that the qualitative shape of $\pi(p)$ will be one of the forms in Figure \ref{fig:prodprofits}.  For certain choices of $w(q)$ (for us, certain values of $a,b$), there are two prices with very comparable profits: one at a low price $p_1$ with a large market share $q(p_1)$, and one at a high price $p_2>p_1$ with a lower market share: $q(p_2)<q(p_1)$.  Moreover, the values of these prices $p_{1,2}$ will depend on the prices of other sellers (even at $\bar J=0$), and which one is optimal can (for some sellers) change based on slight price changes from other sellers.  When these multiple maxima exist, there can be jittery dynamics, \emph{even at $\bar J=0$}, as a consequence of the non-local price updates allowed in our model, and the fact that the competing sellers are not all optimizing the \emph{same} function.

\begin{figure}[t]
    \centering
    \includegraphics[width=\columnwidth]{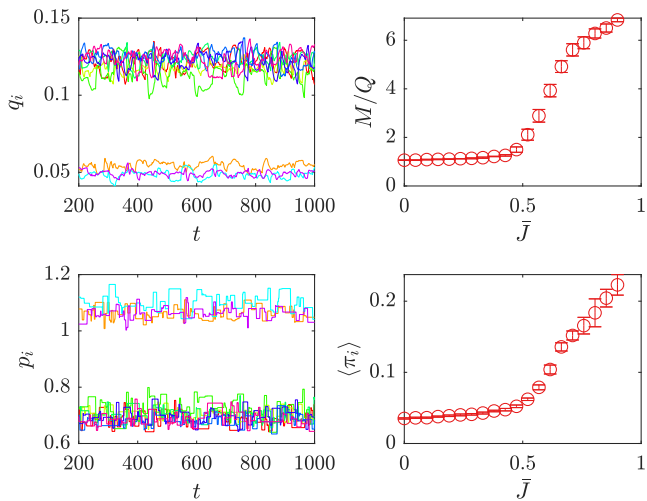}
    \caption{\emph{Left:} Snapshot of $q_i(t)$ and $p_i(t)$ for all $M=10$ sellers of a single run with $\bar J = 0.2$, $a=0.8$, $b=0.5$, $\bar\rho = 1.4$, $\sigma=1$, $\mu=50$.  (Note that $a,b$ match some of the curves in Figure \ref{fig:prodprofits}.)  The market is clearly stable with sellers at two different price points.  \emph{Right:} The PC-NE phase transition at large $\bar \rho$ and increasing $\bar J$ is visible.  Note that, like in the $b=0$ case, the average seller profits $\pi_i$ are ``low" until they abruptly increase in the NE phase.  This emphasizes that sellers at each of the two price points are making low profit in the PC phase.}
    \label{fig:2PC}
\end{figure}

Interestingly, a small amount of network effects (finite $\bar J$) \emph{stabilizes} a ``competitive" phase with two classes of sellers: one at a high price point, and one at a lower price point.  This is \emph{not} a plausible model for the spontaneous emergence of a luxury vs. regular goods market, since recall that in this model on average a buyer is equally likely to prefer any seller's good (assuming all prices are equal).  Rather, this model demonstrates the emergence of two competing seller strategies with somewhat comparable profits.  What network effects do in simulations is ``lock in" sellers to a strategy: since the sellers do not account for network effects, they will interpret the high price goods as having lower intrinsic utility (as the contribution of network effects to $U_{\alpha,i}$ is smaller), and the low price goods as having higher intrinsic utility.  As shown in Figure \ref{fig:2PC}, for small $\bar J$, there are two price points with sellers stably selling at each.   The same figure also demonstrates that when $\bar J > \bar J_{\mathrm{c}}$ for some critical value, there is a transition to the NE phase where both the flip rate $\gamma M$ is finite and O(1), and $Q \ll M$.  In the NE phase, a single monopolist (for a short time) captures a large fraction of all buyers, as in the model with $b=0$.

\begin{figure*}
    \centering
    \includegraphics[width=\textwidth]{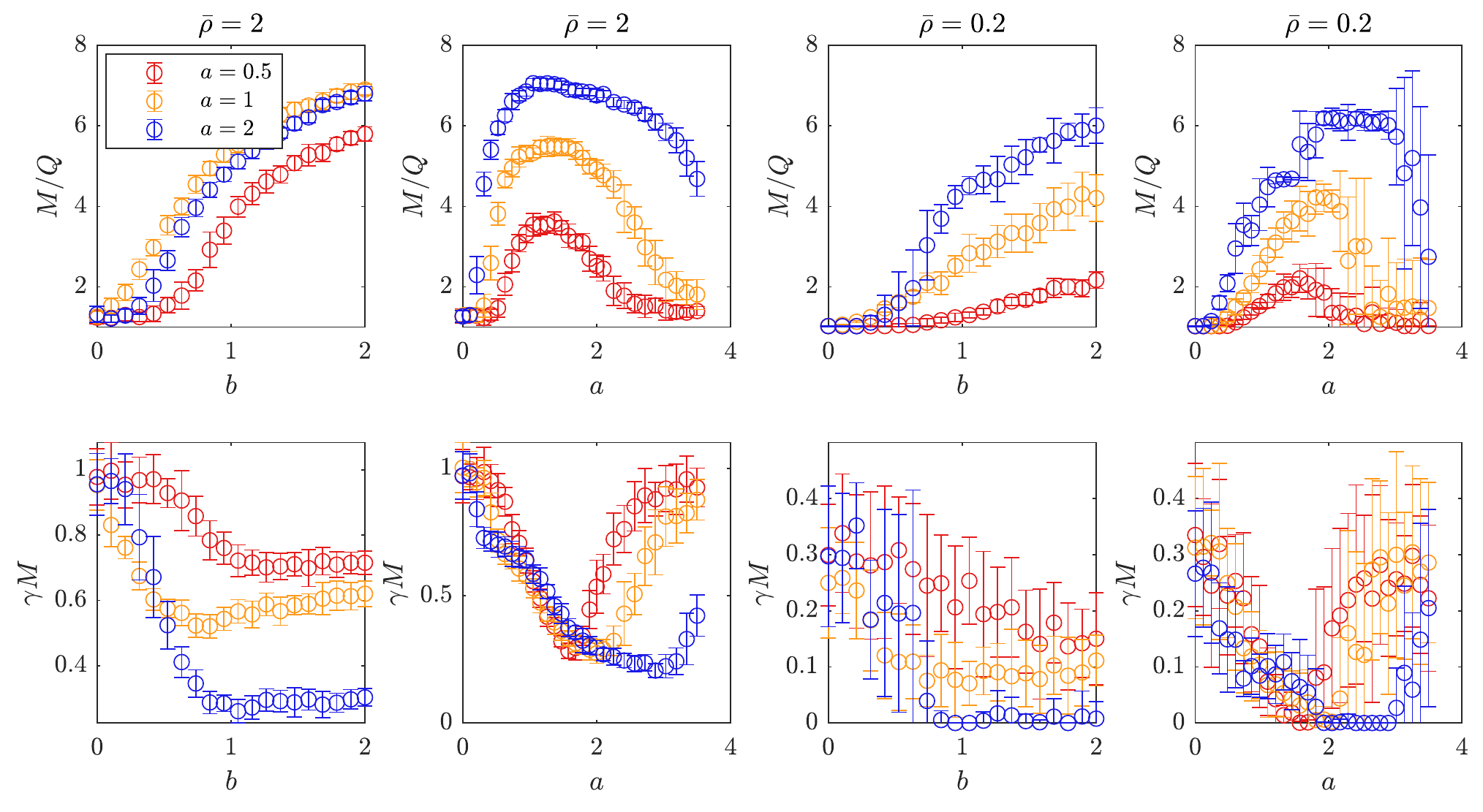}
    \caption{Four panels showing the $a$ or $b$ dependence of $M/Q$ (top row) and $\gamma M$ (bottom row) in simulations with $M=10$, $N=3000$, $\bar J = 0.5$, $\sigma=1$, $\mu=50$.  In the first and third column, we fix $a$ and vary $b$; in the second and fourth columns, we fix $b$ and vary $a$, with colors corresponding to the same values (for either $a$ or $b$, as appropriate) in all figures.  In the left two columns, the buyer dynamics is fast and the naive transition when PC is unstable will be to NE.  This is confirmed as $\gamma M$ stays away from 0, even when $M/Q \gg 1$.   In the right two columns, $\bar \rho$ is reduced by a factor of 10, and we see much more of a tendency to enter the SB (monopolist-dominated) phase.  However, it appears as though the transition is not directly from PC to SB, and there is an intermediate NE phase for a narrow window of parameters.  There are very large sample-to-sample fluctuations in $\gamma$ near the SB-NE transition.  }
    \label{fig:abphases}
\end{figure*}

The effect describes above onsets at a critical value of $b$ (which depends on $a$ -- see Figure \ref{fig:abphases}).  When $b$ becomes particularly large, the tendency towards market instability is simply enhanced even farther, and the PC phase is destabilized.  In contrast, the parameter $a$ should not deviate too strongly from 1 -- if $a\ll 1$ or $a\gg 1$, PC is stabilized.  To understand why, note that in the PC phase each seller has $q_i\approx 1/M$, while only when $q_i=a/M$ does the seller gain an advantage in lower production costs.  When $a$ is too large, sellers will find that it is not worth lowering their price to capture market share $>a/M$ (and thus save on production costs) -- the model will be similar to the one with $b=0$, and with prices shifted $p\rightarrow p+b$ to account for production costs.   Similarly, when $a \ll 1$, all sellers already are taking advantage of negligible production costs even in the PC phase, so the instability out of PC will not be enhanced by much.  These effects are visible in Figure \ref{fig:abphases}.

To summarize, the inclusion of production costs can further destabilize the competitive phase to network effects, but (beyond the emergence of multiple competing seller strategies, possible even without network effects) the overall phenomenology of our model is robust against this effect. 

 \section{Noise}\label{app:noise}
 In this appendix, we discuss the consequences of noise in our model.  There are two natural kinds of noise -- for sellers, and for buyers.
 
 \begin{figure}[t]
    \centering
    \includegraphics[width=\columnwidth]{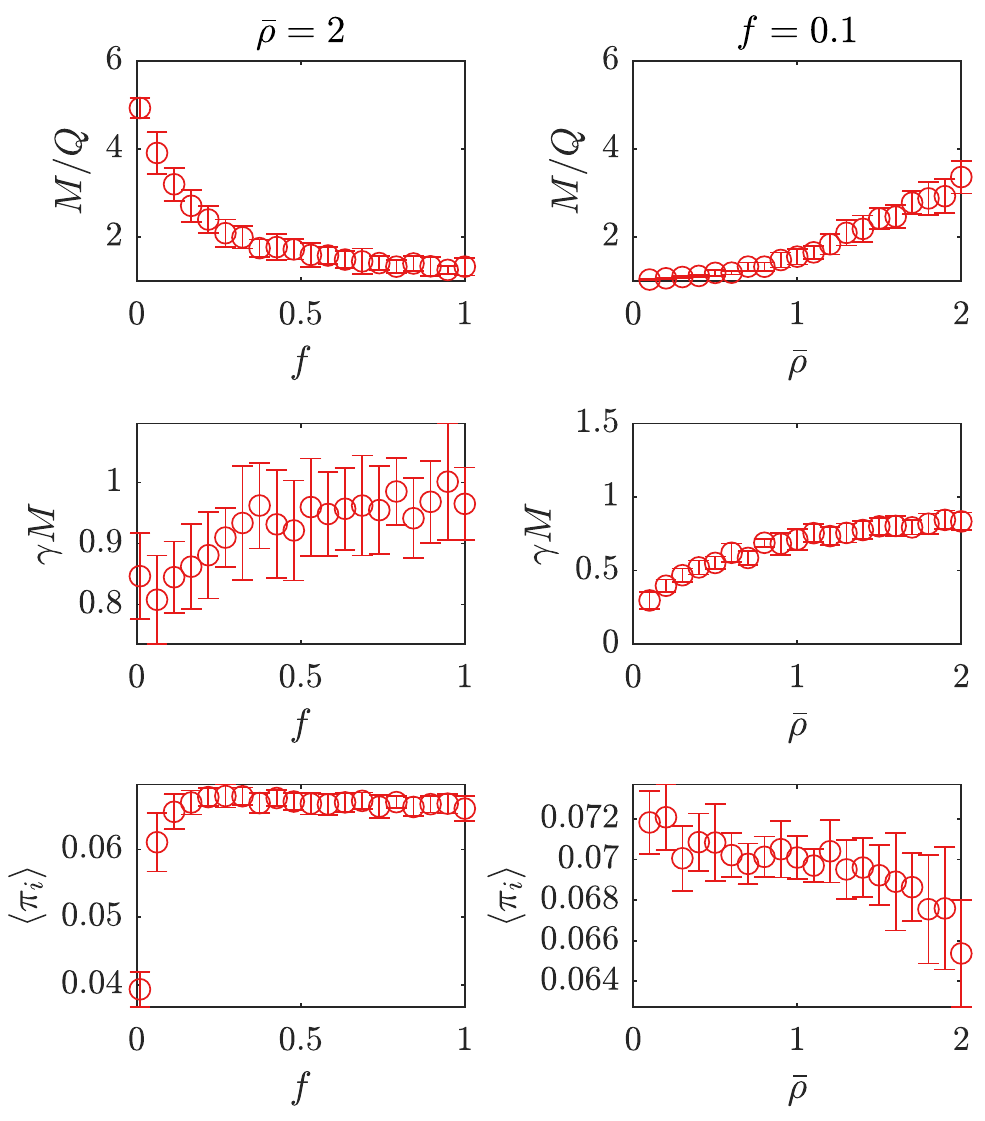}
    \caption{Simulations with $M=10$, $N=3000$, $b=0$, $\sigma=1$, $\mu=50$, $\bar J = 0.5$.  and variable $\bar\rho$ and $f$.   We observe that $M/Q$ sharply decreases upon decreasing $f$, suggesting that decreasing $f$ destabilizes PC.  However, upon decreasing $\bar\rho$, the effects of decreased $f$ are mitigated.}
    \label{fig:sellernoise}
\end{figure}
 
 First, let us discuss ``noise" in the seller dynamics.  Perhaps the most ``realistic" model for noise in the seller dynamics is to simply assume that sellers can only query a small fraction of the buyers before they set their prices (e.g. a focus group or poll only looks at a fraction of the population).  As we have frequently remarked in this work -- the interpretation of these microscopic details is not critical to an  understanding the phase diagram of the model.   Let $f$ denote the fraction of buyers that a seller can poll in any given time step.   A heuristic model for the consequences of $f<1$ is that it effectively enhances finite $N$ effects, by replacing $N \rightarrow f N$.  Therefore, we might expect that many instabilities that, in simulations, appear sensitive to finite $N$ effects, will appear enhanced when $f<1$.  Simulations indeed show that this is largely what happens.  For example, we have argued above that in simulations with small enough $M/N$, finite size effects can destabilize a metastable \cite{leelucas} PC equilibrium below the critical value of $\bar J_{\mathrm{c}}$ (where the PC point becomes strictly unstable) in the large $N$ limit, so we would anticipate that for sufficiently small $f$, we may see the NE phase even when $\bar J < \bar J_{\mathrm{c}}(f=1)$.  This is demonstrated in Figure \ref{fig:sellernoise}.  In the NE phase destabilized by $f<1$ (but not $\bar J > \bar J_{\mathrm{c}}$), seller profits $\langle \pi_i\rangle$ are largely unchanged from their values in the PC phase.  We also find that starting in the NE phase at $f<1$ and $\bar J< \bar J_{\mathrm{c}}$, decreasing $\bar \rho$ does not lead to the SB phase, but rather back to the PC phase.  
 
 There is one qualitative change between the NE phase (with finite $\gamma$ and $M/Q \gg 1$ when $f=1$ versus $f\ll 1$.   In the former, we have seen in the main text that time-averaged seller profits $\langle \pi_i\rangle$ drastically increase in the NE phase.  However, in Figure \ref{fig:sellernoise}, we can see that $\langle \pi_i\rangle$ actually \emph{decreases} with decreasing $f$ -- the market is both strongly fluctuating, yet sellers do not benefit.  This is at least partially explained by the observation that in our model, sellers always make poorer pricing decision with less knowledge (smaller $f$), hence they make smaller profit.  It is interesting that even the condensation of the market onto particular sellers does not counteract this loss of profit due to lower seller information.
 
  \begin{figure*}[t]
    \centering
    \includegraphics[width=\textwidth]{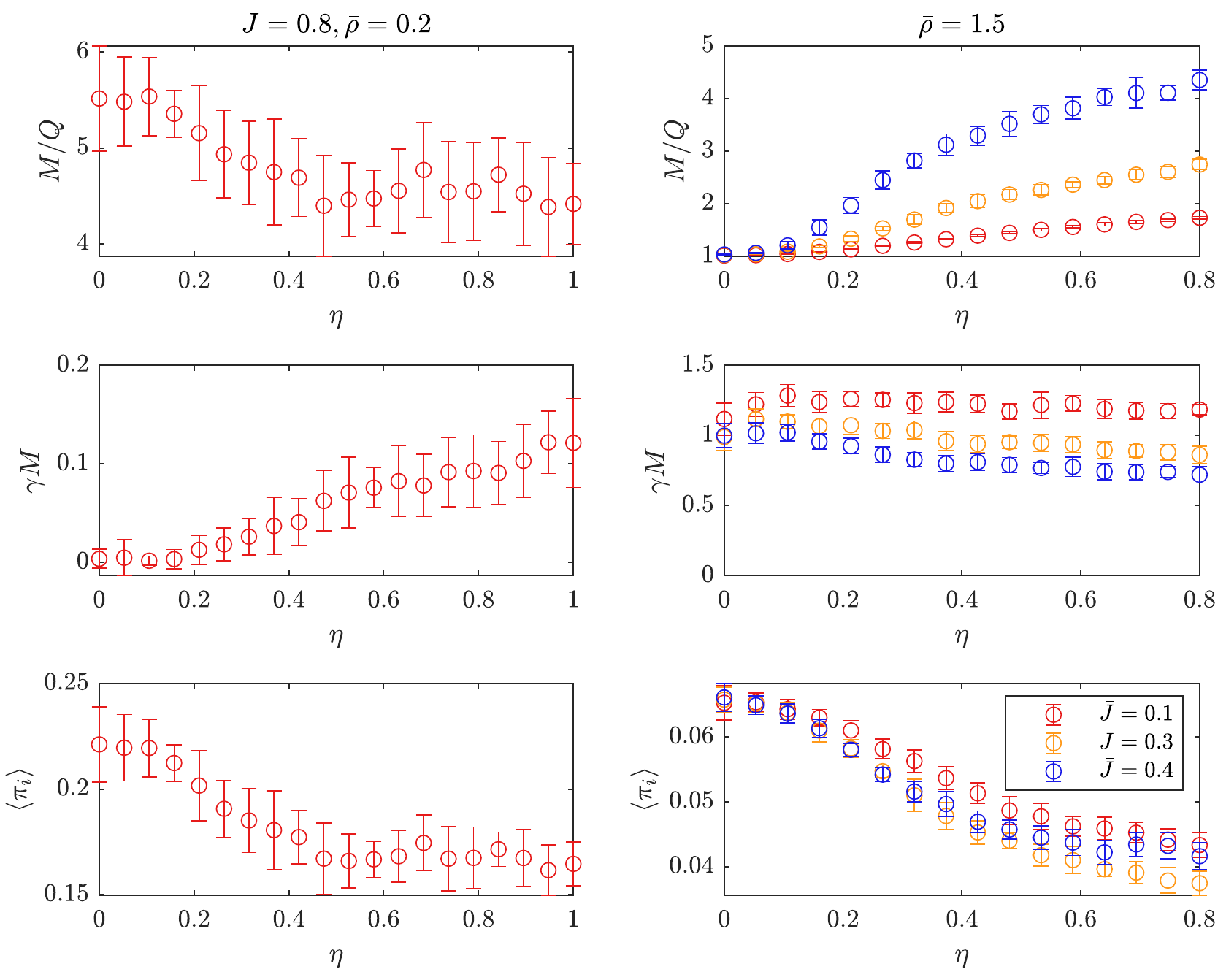}
    \caption{Simulations with $M=10$, $N=3000$, $b=0$, $\sigma=1$, $\mu=50$.  \emph{Left:} The SB phase, characterized by large $M/Q$ and $\gamma \rightarrow 0$, is destabilized by finite $\eta$ and becomes NE (finite $\gamma$).  \emph{Right:} The PC phase also becomes destabilized by finite $\eta$ and leads to an NE phase with finite $\gamma$ and large $M/Q$.  Note that average profit $\langle \pi_i\rangle$ is decreased by finite $\eta$, relative to its value at $\eta=0$. }
    \label{fig:etanoise}
\end{figure*}
 
 An alternative model for seller noise is to instead assume that sellers set their profit-maximizing price not at $p_i^*$ (the argument at which $\pi_i(p_i)$ is optimal), but instead at \begin{equation}
     p_i = p_i^* + \eta Z_t,
 \end{equation}
 where $Z_t$ is a zero-mean unit-variance Gaussian random variable (re-sampled at each time step $t$) -- thus the parameter $\eta$ represents another form of noise.  As shown in Figure \ref{fig:etanoise}, we find that across the phase diagram, increasing $\eta$ largely tends to push the market towards the NE phase -- increasing both the turnover rate $\gamma$ between sellers who have highest market share, and increasing $M/Q$.  This can be understood intuitively as follows.  If we start in the PC phase, because the symmetric stable point is (for moderately large $\bar J$) not the only stable fixed point for the buyer dynamics -- if one seller, for whatever reason, set an unexpectedly low price, they may tilt the market towards a highly asymmetric point where buyers condense onto that seller.  Generally the sellers do not do this if they are profit-maximizing, but in the presence of $\eta$ noise, they may occassionally do this purely due to noise.  Thus we should expect increasing $\eta$ will push PC towards NE.  Instead starting in the symmetry broken phase (SB), where a monopolist has captured the market, we note that increasing $\eta$ makes the monopolist more vulnerable to accidentally overshooting the price that buyers are (with network effects) willing to purchase at.  This will again tend to push the market towards a NE phase.  
 
 In Figure \ref{fig:etanoise}, we observe that the presence of this seller noise again causes the seller profit to slightly decrease -- \emph{even} if the noise pushed the market from PC towards NE.  This is in contrast to the model of the main text, where seller profits always increased dramatically across the PC-NE transition.  This is a qualitative difference between NE phases induced by network effects vs. noisy seller dynamics.
 
  \begin{figure*}[t]
    \centering
    \includegraphics[width=\textwidth]{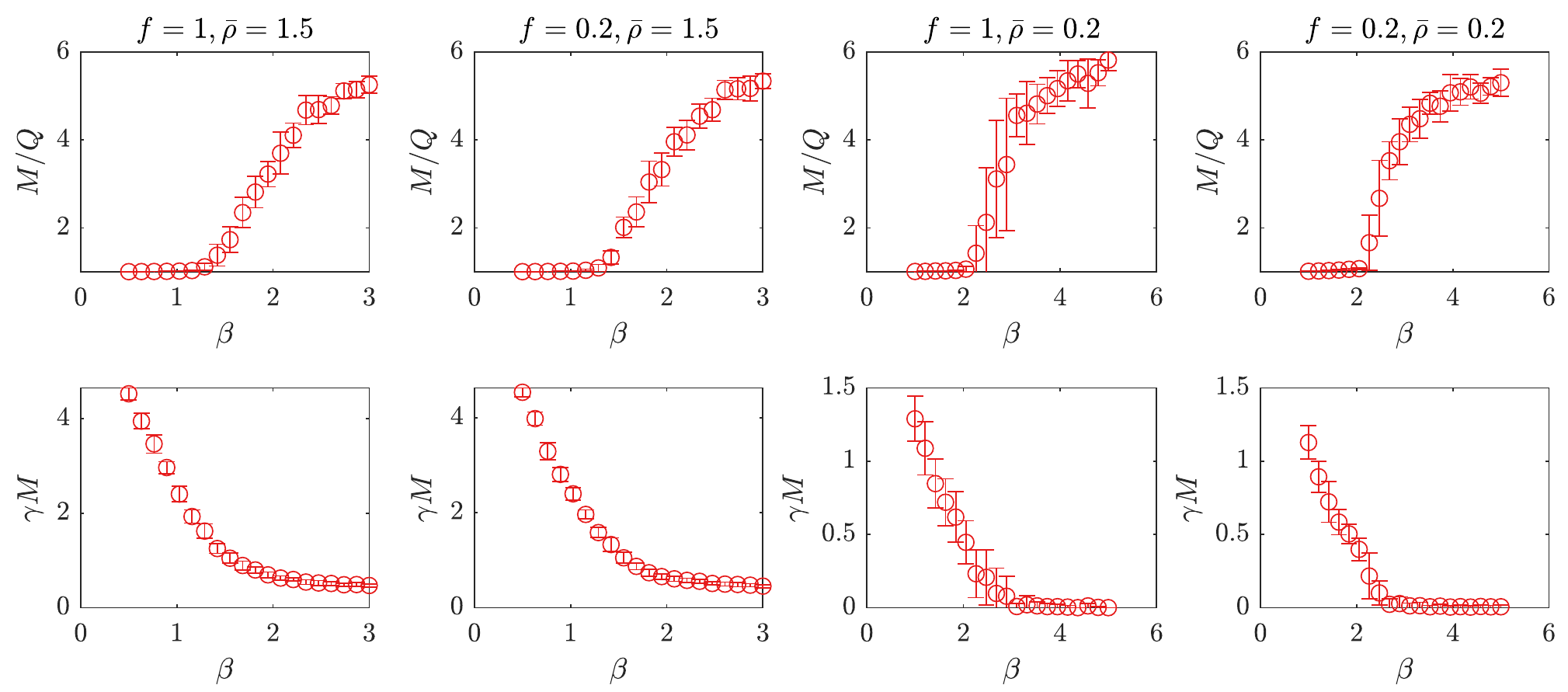}
    \caption{$\gamma M$ and $M/Q$ as a function of $\beta$ for various model runs with $N=3000$, $M=10$, $a=b=0$, $\bar J = 0.85$, $\sigma=1$, $\mu=50$.  Values of $f$ and $\bar \rho$ are written above each column.  In the first and second columns, we observe the PC-NE transition characterized by finite $\gamma$; in the third and fourth columns, we observe the transition from PC to SB ($\gamma \rightarrow 0$ at large $\beta$). There is a narrow window from $2\lesssim\beta\lesssim2.5$ where it appears that the model may be in the NE phase (similar to what is seen in Figure 3): $M/Q$ jumps above 1 \emph{before} $\gamma$ gets extremely close to 0.}
    \label{fig:buyernoise}
\end{figure*}
 
 Second, we discuss noise in the buyer dynamics.  Here, the most natural form of ``noise" corresponds to adding a ``finite temperature", where the probability that a buyer picks choice $i$ is given by \begin{equation}
 \mathbb{P}[x_i = n] = \frac{\mathrm{e}^{\beta U_{\alpha,n}}}{\sum_{m=0}^M \mathrm{e}^{\beta U_{\alpha,m}}}.    
 \end{equation}
 Here $\beta$ is the effective inverse temperature, and each buyer tries to minimize effective energy $-U_{\alpha,i}$ among options $i$ (with some fluctuations).  The model in the main text corresponds to $\beta=\infty$.  In statistical physics, turning on such thermal fluctuations tends to promote ``disordered" phases, where all symmetries are restored in the thermodynamic ($N\rightarrow \infty$) limit.  Therefore, we expect that decreasing the value of $\beta$ will (for small enough $\beta$) always lead to the PC phase.  Figure \ref{fig:buyernoise} shows that this is indeed the case.   This same figure also demonstrates that when $f<1$, decreasing $\beta$ again enhances the PC phase.
 
 The most important conclusion from this appendix is that the qualitative findings of our model are robust to various ``imperfections" or noise in the economic dynamics.  Such noise is to be expected given realistic conditions of imperfect information, but it does not lead to any major changes in our model.  In particular, the key conclusion of this paper -- that strong network effects can cause persistent dynamics in a market where buyers and sellers maximize only immediate-term utility/profit respectively -- is not a finely-tuned result.
 
  \section{On $P(q_i)$}\label{app:pqi}
  
\begin{figure}
 \includegraphics[width=\columnwidth]{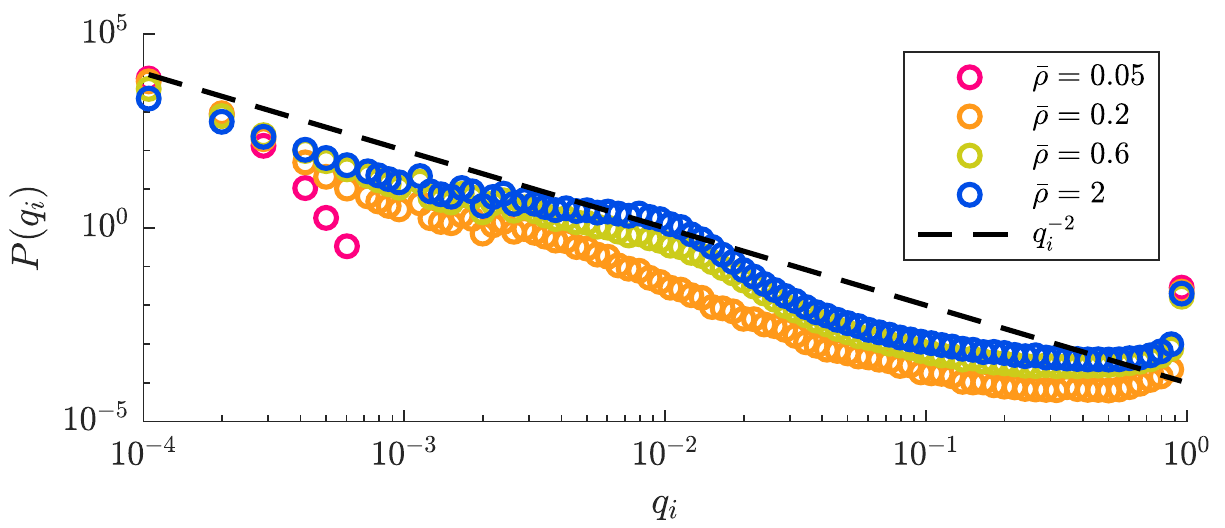}
 \includegraphics[width=\columnwidth]{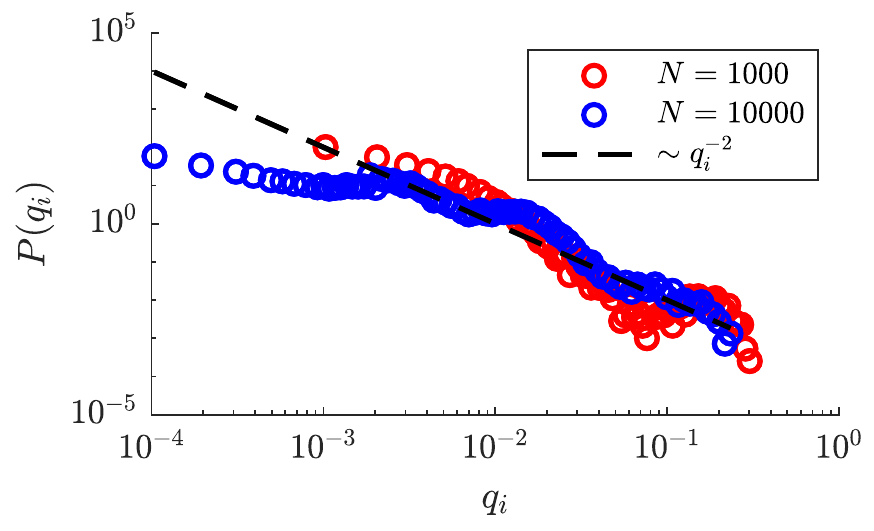}
 \caption{\emph{Top:} Simulations of the time-averaged $P(q_i)$ with $M=100$, $N=100M$, $\bar J = 0.9$, $\sigma=1$ and $\mu=5$.  \emph{Bottom:} Simulations of $P(q_i)$ in the NE phase, averaged only over times at which the seller with the largest market share has just flipped, with $M=100$, $\bar J = 0.7$, $\bar\rho = 0.9$.}
 \label{fig:Pqi}
 \end{figure}

 Figure \ref{fig:Pqi} shows the $\bar\rho$ dependence of the distribution $P(q_i)$, which demonstrates that the distribution seems to have the best power law scaling close to the NE-SB transition when $\bar \rho \sim \frac{1}{3}$.
 It also shows that at the times where (in the NE phase) the seller with the largest market share changes, we can already see broad distributions in $q_i$, implying that the heavy tailed distribution does not come \emph{entirely} from the time average (though this certainly does ``smooth out" the distribution further).

 We remark that in Figure \ref{fig:loglogqt} the heterogeneous nature of the distribution of firm sizes is visible.  At any fixed time, as remarked above, we do not see 2 decades of power-law scaling in the distribution of $P(q_i)$ (although heterogeneous distributions are present): the somewhat more smooth scaling seen in Figure 4 seems to arise in part due to time-averaging.  An analytic understanding of the distribution of $q_i$ as a function of time would be valuable.
 
 \section{Distinguishing the NE phase from exogenous shocks}\label{app:shock}
In an ideal setting (or perhaps in an artificial market), the existence of the NE phase would be best studied by comparing the dynamics in a market where buyers do not communicate with each other, versus one where they do.  In a real market, we cannot arbitrarily tune network effects.  How can we check whether we are in the NE phase?  Answering this question is substantially complicated by the fact that the real world is not in ``equilibrium".  There \emph{are} market shocks, technological innovations, etc.   Therefore, simply finding prices that fluctuate with time is not enough to show that the model of this paper is relevant for a real market -- one must also argue that the origin of these fluctuations is endogenous (intrinsic to the market), not exogenous (caused by external ``driving", such as abrupt price changes in an adjacent market).

We propose, as sketched in the main text, comparing the ``forward" vs. ``backwards" time correlations as a sensible check for whether the fluctuations in a market are endogenous or exogenous.  In a nutshell, we will argue that the fastest fluctuations in an exogenously-driven market are when the market undergoes condensation ($Q$ decreasing), but that in our model it is opposite -- the fastest fluctuations are when the market is returning to a ``permutation symmetric" state ($Q$ increasing).  Intuitively, the exogenous shocks that do not uniformly affect the entire market will be abrupt in time, causing rapid adjustments that the market must slowly adapt to -- the largest jump in $Q$ will occur right at the location of the shock.  Furthermore, we expect $Q$ will tend to \emph{decrease} just after a shock, since the system will try to relax to a competitive equilibrium $Q\rightarrow 1$ if network effects are not responsible for the non-equilibrium dynamics (and, as in our model, all sellers are on average producing goods of the same quality).  However, in the NE phase, as discussed in the main text, there is an instability (cascading avalanche of decision changes) that precedes the condensation of the market onto a single seller.  This process is not instantaneous.  However, when a seller overprices and condensation ends, the collapse in monopolist market share will be ``unstoppable" if buyer dynamics is too fast -- this is the fastest dynamics in the problem, so the most rapid changes in $Q(t)$ will be when it increases back to $M$.   This effect is also most pronounced closer to the NE-PC transition (smaller $\bar J$): here the instability has a slower rate (so condensation is slower), while the decline in a monopolist's market share after overpricing occurs at a roughly $\bar J$-independent rate.

  A crude, but easily calculable ``correlation function" is $R(t)$, defined in a few steps as follows.   First, let us define \begin{equation}
C(t) = \frac{1}{N_t-t}\sum_{s=1}^{N_t-t} \min\left(0,\frac{M}{Q(s+t)} - \frac{M}{Q(s)}\right)^2. \label{eq:Ct}
\end{equation}
where $N_t$ is the number of time steps in the simulation.   $C(t)$ is a type of correlation function, although it is not of the usual kind studied in physics, which would be of the form $\langle \frac{M}{Q}(t+s)\frac{M}{Q}(s)\rangle$. The reason we do not study these more conventional correlators is simple: this statistical correlation function is by construction is even in $t$, and (as stressed above) our test for exogenous vs. endogenous dynamics will be sensitive to the sign of $t$.   The max in (\ref{eq:Ct}) counts only the times when $\frac{M}{Q}(s+t)>\frac{M}{Q}(s)$.   We also square $(\frac{M}{Q}(s+t)-\frac{M}{Q}(s))^2$ to avoid the following problem with the linear function: since on long time scales $Q$ does not increase or decrease, the rate of increase/decrease will be 0, meaning again the $t>0$ and $t<0$ functions will approach the same value.   By squaring the function, we count more the time steps when $Q(s+t)$ jumps by a lot relative to $Q(s)$:  if $t>0$ and $\frac{M}{Q}(s+t)>\frac{M}{Q}(s)$, this big jump will contribute a lot to $C(t)$, while if $\frac{M}{Q}(s+t)<\frac{M}{Q}(s)$, the jump will contribute instead to $C(-t)$ (if $s$ is not too close to the start/end of the time-sampling period!).  We thus look at \begin{equation}
R(t) = \frac{C(-t)}{C(t)}
\end{equation}
for $t\ge 1$.  If $R<1$, it means that the largest magnitude jumps are associated with $M/Q$ increasing; if $R>1$, they correspond to $M/Q$ decreasing.   Hence we expect a model of exogenous shocks will have $R<1$ at small $t$, while in our model we have argued above that $R>1$.

  \begin{figure}[t]
    \centering
    \includegraphics[width=\columnwidth]{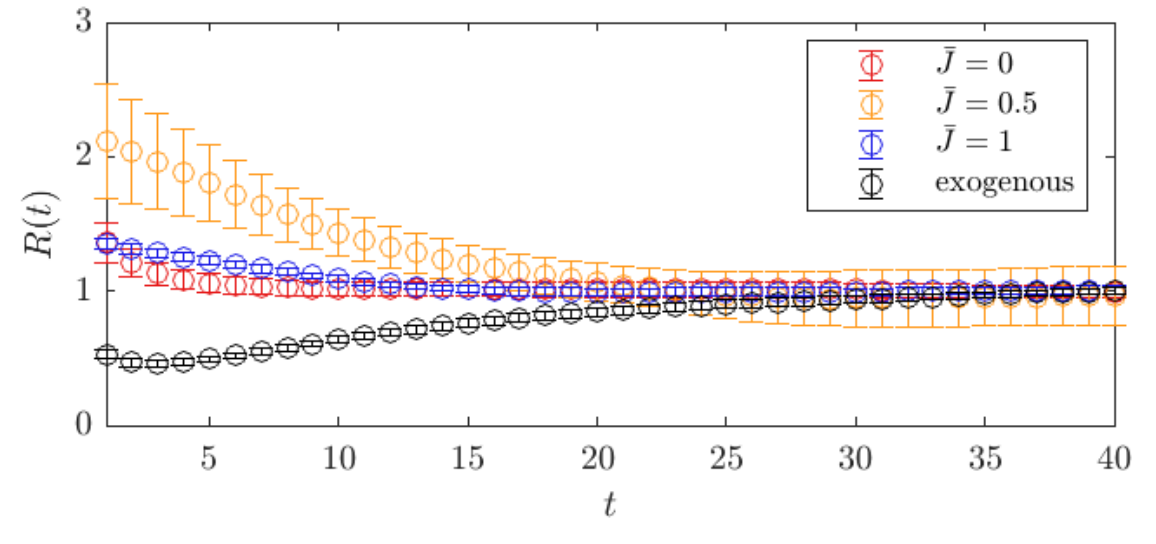}
    \caption{Checking $R(t)$ in simulations with $M=10$, $N=3000$, $\sigma=1$, $\mu=50$, $\bar\rho=1.5$ (and no production costs or noise).   For these parameters, the NE phase arises at $0.6\lesssim \bar J_{\mathrm{c}} \lesssim 0.7$.   We see that $R(t)>1$ and is largest for small $t$ near the PC-NE transition.   In contrast, for the model of exogenous shocks described in the text, $R(t)<1$ at small $t$, implying this ``correlation function" is sensitive to whether fluctuations are driven by network effects or not.}
    \label{fig:Rt}
\end{figure}

Figure \ref{fig:Rt} confirms these expectations about our model.    Interestingly, even at $\bar J=0$, there appears to be a modest tendency for $R>1$ at very short times.  As expected, the largest increases in $R$ arise near the NE-PC transition.   We use the following cartoon to model a market where the largest fluctuations are driven by exogenous shocks, but without network effects ($\bar J=0$):  every $7M$ time steps (this precise number is unimportant, although we choose it to large compared to $M$ so sellers have time to respond to the shock), with 50\% probability we increase each seller's price by $2\sigma$ (which will make a huge fraction of buyers choose another seller).  As expected, this model is characterized by predictable bursts where $Q$ decreases every $7M$ time steps, and $R<1$ for $t<7M/2$.  Since the behavior of $R(t)$ on relatively short times is of most interest, we deduce that our proposed ``correlation function" $R(t)$ indeed distinguishes the origin of temporal fluctuations in and near the NE phase from a model of exogenous shocks.

\end{appendix}

\bibliography{thebib}

\end{document}